\documentclass[usenatbib,usegraphicx]{mn2e}
\usepackage{amsfonts}
\usepackage{amsmath}
\usepackage{amssymb}
\usepackage{comment}
\usepackage{color}
\usepackage{url}
\usepackage{morefloats}

\title[\textbf{Measuring galaxy morphology at $z>1$}]
    {\textbf{Measuring galaxy morphology at $z>1$. I - calibration of automated proxies}}

\author[Huertas-Company et al.]
  {M.~Huertas-Company$^{1}$,
  S.~Kaviraj$^{2}$, S.~Mei$^{1}$, R.W.~O'Connell$^3$, R.A.~Windhorst$^4$
   \newauthor
  S.H.~Cohen$^4$, N.P~Hathi$^5$, A.M.~Koekemoer$^6$, R.~Licitra$^1$, A.~Raichoor$^1$, M.J.~Rutkowski$^7$ \\
$^{1}$University Denis Diderot, CNRS, GEPI-Observatoire de Paris UMR 8111, Paris, France\\
$^{2}$Centre for Astrophysics Research, University of Hertfordshire, College Lane, Hatfield, Herts AL10 9AB, UK\\
$^{3}$Astronomy Department, University of Virginia, P.O. Box 3818, Charlottesville, VA 22903, USA\\
$^{4}$School of Earth and Space Exploration, Arizona State University, Tempe, AZ 85287-1504, USA\\
$^{5}$Aix Marseille Universit\'{e}, CNRS, LAM (Laboratoire d'Astrophysique de Marseille) UMR 7326, 13388, Marseille, France\\
$^{6}$Space Telescope Science Institute, 3700 San Martin Drive, Baltimore, MD 21218, USA\\
$^{7}$University of Minnesota, Minneapolis, MN 55405 USA}

\begin{document}

\maketitle

%%%%%%%%%%%%%%%%%%%%%%%%%%%%
% symbols for references:  %
%%%%%%%%%%%%%%%%%%%%%%%%%%%%
\def \aj {AJ}
\def \mnras {MNRAS}
\def \pasp {PASP}
\def \apj {ApJ}
\def \apjs {ApJS}
\def \apjl {ApJL}
\def \aap {A\&A}
\def \nat {Nature}
\def \araa {ARAA}
\def \iaucirc {IAUC}
\def \aaps {A\&A Suppl.}
\def \qjras {QJRAS}
\def \na {New Astronomy}
\def \aapr {A\&ARv}
\def\lesssim{\mathrel{\hbox{\rlap{\hbox{\lower4pt\hbox{$\sim$}}}\hbox{$<$}}}}
\def\gtrsim{\mathrel{\hbox{\rlap{\hbox{\lower4pt\hbox{$\sim$}}}\hbox{$>$}}}}

%.........................................................................................................

\begin{abstract}
New near-infrared surveys, using the HST, offer an unprecedented
opportunity to study rest-frame optical galaxy morphologies at
$z>1$ -- a critical epoch when most of today's massive galaxies
formed -- and to calibrate automated morphological parameters that
will play a key role in classifying future massive datasets like
EUCLID or LSST. We study automated parameters (e.g. CAS, Gini, $M_{20}$)
of massive ($\log[M_*]>10.5$) galaxies at $1<z<3$, measure their
dependence on wavelength and evolution with redshift and quantify
the reliability of these parameters in discriminating between
visually-determined morphologies, using a Support Vector Machine
based method. We find that the relative trends between
morphological types observed in the low-redshift literature are
preserved at $z>1$: bulge-dominated systems have systematically
higher concentration and Gini coefficients and are less asymmetric
and rounder than disk-dominated galaxies. However, at $z>1$,
galaxies are, on average, $\sim50\%$ more asymmetric and have Gini
and $M_{20}$ values that are $\sim10\%$ higher and $\sim20\%$
lower respectively (the differences being more pronounced for
bulge-dominated systems). In bulge-dominated galaxies,
morphological parameters derived from the rest-frame UV and
optical wavelengths are well correlated; however late-type
galaxies exhibit ($40-60\%$) higher asymmetry and clumpiness when
measured in the rest-frame UV. We find that broad morphological
classes (e.g. bulge vs. disk dominated) can be distinguished using
parameters with high (80\%) purity and completeness of $\sim80\%$.
In a similar vein, irregular disks and mergers can also be
distinguished from bulges and regular disks with a contamination
lower than $\sim20\%$. However, mergers cannot
be differentiated from the irregular morphological class using these
parameters, due to increasingly asymmetry of non-interacting
late-type galaxies at $z>1$. Our automated procedure is applied to the CANDELS GOODS-S field and compared with the visual classification recently released on the same field getting similar results. The effects of performing the classification in different filters simultaneously are also discussed.

%Using our calibrated classifier we provide an automated morphological classification of CANDELS-GOODS freely available upon request.

\end{abstract}

%.........................................................................................................


%.........................................................................................................

\section{Introduction}
Our currently-accepted $\Lambda$CDM paradigm of structure
formation postulates a hierarchical growth of mass, with smaller
systems merging, under the influence of gravity, to form
progressively larger ones
\citep[e.g.][]{White1978,Cole2000,2003MNRAS.341...54K,Hatton2003,Somerville2012}.
As `endpoints' of this hierarchical process, massive galaxies
dominate the stellar mass density at the present day
\citep[e.g.][]{2014MNRAS.437L..41K}, making them key laboratories for
studying the evolving Universe over cosmic time. The literature
indicates that the bulk of the stars in today's massive galaxies
are old and formed at $z>1$
\citep[e.g.][]{Ellis97,Stanford98,Trager2000a,Bernardi2003,Bell2004,Kaviraj2005,Faber2007,Kaviraj2008,2012ApJS..199....4R}.
Understanding the processes that drove the buildup of this
old stellar mass is therefore a central topic in observational
cosmology.

Morphology carries a unique imprint of the processes that are
currently driving star-formation in a galaxy. For example, the
presence of multiple nuclei and extended tidal features indicates
that some fraction of the star-formation is likely to be
driven by an ongoing merger. Equally, the absence of such
features implies that the evolution of the galaxy is likely to be
dominated by internal (`secular') processes. Thus, reliable
determinations of galaxy morphologies at high redshift are
critical for understanding the processes that drove early star
formation and created the stars that dominate the Universe today.

Measurement of galaxy morphology, especially at high redshift
where systems appear less extended, demands high-resolution
imaging, such as that offered by the \emph{Hubble Space Telescope}
(HST). In addition, morphology is best determined in the
\emph{rest-frame} optical wavelengths, which trace the bulk,
underlying stellar population of the galaxy (e.g \citealp{2005ApJ...623..721P, 2012ApJS..199....4R}). Measurement of
morphology at $z>1$ has traditionally been challenging, as most
HST surveys at high redshift have exploited optical filters that
trace rest-frame ultraviolet (UV) wavelengths at these epochs.
However, since the UV is heavily dominated by young stars
\citep{Martin2005,Kaviraj2007a}, the UV morphology largely traces
the star-forming regions, which may not be representative of the
underlying older stellar population of the galaxy (unless the system is
formed exclusively of young stars).

Following the GOODS-NICMOS survey \citep{2011MNRAS.413...80C}, a new
generation of surveys, using near-infrared (NIR) filters on HST's Wide
Field Camera 3 (WFC3), are providing unprecedented access to the
rest-frame optical morphologies of galaxies at $z>1$. The first
such observing campaign was a 45 arcmin$^2$ survey of the
GOODS-South field using the WFC3's $Y$, $J$ and $H$-band filters,
as part of the WFC3 Early-Release Science (ERS) programme (Windhost et al. 2011). This
has been followed by CANDELS - the largest HST Treasury programme
to date \citep{Grogin2011,Koekemoer2011} - which has imaged 800
arcmin$^2$ of the sky.
Morphological studies using these new datasets will be an
important feature of galaxy evolution work in the forthcoming
literature.

A variety of methods have traditionally been used to perform
morphological classification. Quantitative parameters, such as
Concentration (C), Asymmetry (A), Clumpiness (S), M$_{20}$ and the
Gini coefficient, have often been employed as morphological
proxies, particularly for studying large survey datasets
\citep[see
e.g.][]{Abraham1996,Abraham2003,Conselice2003,Lotz2004,Lotz2008,Huertas-Company2008,Huertas-Company2009,Huertas-Company2011}.
Methods like galSVM \citep{Huertas-Company2008,
Huertas-Company2011}, that makes use of learning machines and
improved data mining techniques specially adapted to large
surveys, are able to leverage the power of these morphological
parameters to quantify the probability of a galaxy belonging to a
certain morphological class. Studies using morphological
parameters are typically tested and calibrated on visual
classifications \citep[e.g.][]{1996ApJ...472L..13O, 2004MNRAS.348.1038B, Huertas-Company2011},
which is arguably the most accurate method of quantifying galaxy
morphology.

However, visual inspection can be prohibitively time-consuming for
individuals (or small groups of researchers) when large volumes of
data are involved, making it somewhat impractical for
morphological work using modern surveys, although small subsamples
of modern survey datasets have been classified via this technique
\citep[e.g.][]{Bundy2005,2007AJ....134..579F,Jogee2009,Nair2010,2012MNRAS.423...49K}.
A potential solution is to dramatically increase the
number of people involved in performing visual classifications,
using projects like Galaxy Zoo
\citep[][GZ;]{2008MNRAS.389.1179L, 2011MNRAS.410..166L}. GZ has used more than
800,000 members of the general public to morphologically classify,
via visual inspection, the entire SDSS spectroscopic sample
($\sim$1 million galaxies). Subsequent incarnations of the project
have classified, or are currently classifying, HST surveys
including programmes like CANDELS. While GZ's crowd sourcing
capability has enabled visual classifications of large survey
datasets, even projects such as this do not have the capacity to
process the prodigious amounts of data that are expected
from the next generation of surveys. For example, GZ would require
more than a hundred years to classify all data from the
forthcoming LSST and EUCLID missions with its current
user capacity. Thus, for future datasets, efficient and
well-tested algorithms, trained on visual classifications is
likely to be the only approach for effectively studying galaxy
morphology. Given the focus of future datasets on the early
Universe, it is desirable and timely to explore how well different
morphological classes at $z>1$ can be identified using
morphological parameters and if visual classifications can be used
to train intelligent algorithms. This is the main goal of this
paper.

In this study, we use a sample of bright ($H<24$) galaxies at
$1<z<3$, drawn from the WFC3 ERS and CANDELS programmes, to explore the
performance of parameters in separating galaxies in different
morphological classes. {While some recent work has started
addressing this problem \citep[e.g.][]{2013ApJ...774...47L, 2013MNRAS.433.1185M}, no concise
quantification of the accuracy of automated classifications exists
to date. This paper is organized as follows. In Section
\ref{sec:data}, we outline the properties of the ERS sample that
underpins this study and describe the morphological parameters
computed in this work. In Section~\ref{sec:props}, we quantify the
properties of the parameters, their evolution with redshift and
their wavelength dependence. Finally, in
Section~\ref{sec:auto_morph} we explore the ability of sets of
morphological parameters to determine galaxy morphologies at $z>1$
with the SVM based code galSVM. We quantify the
completeness and purity of the samples obtained, analyze the
effect of the specific visual training used, and the impact of the
number of parameters employed. Section~\ref{sec:candels}, explores how the automatic classification compares to the CANDELS visual classification \citep{2014arXiv1401.2455K}. Finally, in section~\ref{sec:multi_lambda}, we investigate the possibility of performing morphological classifications in different filters simultaneously. We summarize our findings in
Section~\ref{sec:summary}. Throughout, we use the WMAP7
cosmological parameters \citep{2011ApJS..192...18K} and present
photometry in the AB magnitude system \citep{Oke1983}.

%.........................................................................................................

\section{Data}
\label{sec:data}
\subsection{High-redshift galaxy sample}

\subsubsection{WFC3-ERS data}
The WFC3 ERS programme has imaged around one-third of the GOODS-South field
with both the UVIS and IR channels of the WFC3. The observations, data
reduction, and instrument performance are described in detail in Windhorst
et al. (2011) and summarised here. The goal of this part of the ERS
programme was to demonstrate WFC3's capabilities for studying high-redshift
galaxies in the UV and NIR, by observing a portion of GOODS-South
\citep{Giavalisco2004}. The total WFC3 exposure time was 104 orbits, with 40
orbits in the UVIS channel and 60 orbits of  NIR imaging. The UVIS data
covered $\sim$55 arcmin$^2$, in each of the F225W, F275W and F336W filters,
with relative exposure times of 2:2:1. The IR data covered $\sim$45
arcmin$^2$ using the F098M ($Y_s$), F125W ($J$), and F160W ($H$) filters
with equal exposure times of 2 orbits per filter. The data were
astrometrically aligned with version 2.0\footnote{http://archive.stsci.edu/pub/hlsp/goods/v2/} of the GOODS-S HST/ACS data
Giavalisco et
al. 2004), that was rebinned to have a pixel scale of $0\farcs090$ per pixel.
Together, the WFC3 ERS data provide 10-band HST panchromatic coverage over 0.2 - 1.7
$\mu$m, with 5$\sigma$ point-source depths of $AB \sim 26.4$ mag, and $AB
\sim27.5$ mag in the UV and IR, respectively.

\subsubsection{Sample selection}
In the bulk this paper we study 628 ERS galaxies that are brighter than $H=24$ mag and
have either spectroscopic or photometric redshifts in the range $1<z<3$. For section~\ref{sec:candels}, we will consider an enlarged CANDELS sample of $\sim 1800$ galaxies from the morphology catalog by \cite{2014arXiv1401.2455K}.
Photometric redshifts were calculated by applying the EAZY code
\citep{Brammer2008} within the ERS collaboration from the 10-band WFC3/ACS photometric catalogue.
Spectroscopic redshifts are drawn from the literature, from spectra taken
using the Very Large Telescope
\citep{2004A&A...428.1043L,2004ApJS..155..271S,2005A&A...437..883M,2007A&A...465.1099R, 2008A&A...478...83V, 2009A&A...494..443P},
the Keck telescopes \citep{2004ApJ...613..200S} and the HST ACS grism
\citep{2005ApJ...626..680D,2006ApJ...636..115P,2009ApJ...706..158F}. For the analysis that follows, spectroscopic redshifts are always used where
available.

\subsubsection{Morphological parameters}
\label{sec:morpho_params} In addition to the photometry,
stellar masses, photometric redshifts provided in the ERS
catalog (Seth et al., in prep.), we measure widely used morphological parameters
for all galaxies in our sample.

\begin{itemize}

\item {\bf Concentration: }Concentration \emph{(C)} measures the ratio of light within a circular or
elliptical inner aperture to that in an outer aperture. Here, we adopt the
definition of $C$ in \cite{Bershady2000} i.e. the ratio of the circular
radii containing 20\% and 80\% of the total flux. In other words: $C = 5
\log{(r_{80}/r_{20})}$. Following \cite{Conselice2003}, the total flux is
defined as the flux contained within 1.5 $r_P$, where $r_P$ is the Petrosian
radius. The galaxy centre is determined by minimising the asymmetry of the
image (as described below).

%.........................................................................................................

\item {\bf Asymmetry:} Asymmetry ($A$), a measure of the degree to which the galaxy light is
rotationally symmetric, is calculated by subtracting the galaxy image
rotated by 180 degrees from the original image \citep{Conselice2003}. Thus:

\begin{equation}
A = \frac{1}{2} \left( \normalsize \frac{\sum |I(i,j)-I_{180}(i,j)|}{\sum
I(i,j)}-\frac{\sum |B(i,j)-B_{180}(i,j)|}{\sum I(i,j)} \right),
\end{equation}

where $I$ is the galaxy image, $I_{180}$ is the galaxy image rotated by 180
degrees about the galaxy's central pixel and $B$ is the average asymmetry of
the background. The central pixel is determined by minimizing $A$ over (x,y).

%.........................................................................................................

\item {\bf Smoothness/Clumpiness:} Smoothness ($S$) or Clumpiness, a measure of the degree of small-scale structure in a
galaxy \citep{Conselice2000}, is calculated by smoothing the galaxy image by
a boxcar of a given width and then subtracting this from the original image.
Thus:

\begin{equation}
S = \frac{1}{2} \left( \normalsize \frac{\sum |I(i,j)-I_{S}(i,j)|}{\sum
I(i,j)}-\frac{\sum |B(i,j)-B_{S}(i,j)|}{\sum I(i,j)} \right),
\end{equation}

where $I_S$ is the galaxy image and $B_S$ is the background, both smoothed
by a boxcar of width 0.25 $r_P$.

%.........................................................................................................

\item {\bf M$_{20}$: } The total second-order moment M$_{tot}$ is the flux in each pixel ($f_i$)
multiplied by the square of the distance to the galaxy centre, summed over
all galaxy pixels. In other words, M$_{tot}$ = $\sum f_i
[(x_i-x_c)^2+(y_i-y_c)^2]$, where ($x_c$,$y_c$) is the galaxy centre. The
second-order moment of the brightest regions of the galaxy traces the
spatial distribution of bright nuclei, bars, spiral arms and off-centre star
clusters. Following \citet{Lotz2004}, we define M$_{20}$ as the normalized
second-order moment of the 20\% brightest pixels of the galaxy. The logarithm of that value is used throughout the paper.

%.........................................................................................................

\item {\bf Gini coefficient:}
The Gini coefficient ($G$) is a rank-ordered cumulative distribution
function of a population's "wealth", or in this case a galaxy's pixel values
\citep{Abraham2003}. In most local galaxies, $G$ correlates with $C$ and
increases with the fraction of light in the central regions. However, unlike
$C$, $G$ is independent of the large-scale spatial distribution of the
galaxys light. Thus, $G$ differs from $C$ in that it can distinguish
between galaxies with shallow light profiles (which have both low $C$ and
$G$) and galaxies where most of the flux is located in a few pixels but not
at the centre (having low $C$ but high $G$).

%.........................................................................................................

\item {\bf Sersic index: } Finally, we use an estimate of
the Sersic index ($n$) measured using the WFC3 F160W images. These
are calculated using GALAPAGOS (Barden et al. 2012), an IDL based
pipeline to run SEXTRACTOR (Bertin \& Arnouts 1996) and GALFIT
(Peng et al. 2002) together. Individual galaxies are fitted with a
2D Sersic profile, using the default GALAPAGOS parameters
(see Haussler et al. 2007).

\item {\bf Ellipticity:} We will also use as input parameter to discriminate between different morphologies the ellipticity ($e=1-b/a$) as measured by Sextractor (i.e. without deconvolution) which measures the ratio between the major and minor axis of the galaxy.

\end{itemize}

%.........................................................................................................

\subsubsection{Visual morphologies}
\label{sec:visual} Finally, we perform a visual
classification of all galaxies using the H band filter (F160W).
Galaxies are classified into 6 morphological classes: (1) bulges,
(2) bulges with faint disks, (3) regular disks, (4) irregular
disks (which includes clumpy disks), (5) mergers and (6)
unclassifiable systems. Bulges are systems that appear symmetric
and concentrated with no apparent sign of disk structure. Bulges
with faint disks include all galaxies which are clearly bulge
dominated but exhibit a faint disk around the bulge. Regular disks
are the equivalent of normal spiral galaxies in the local
universe, i.e systems that have a significant disk, with a small
or insignificant bulge in the centre. In the irregular disk class,
we include all galaxies which are clearly disk dominated but
present significant asymmetries (e.g. clumps) or disturbances which
make them unlike a conventional spiral galaxy. The merger class
contains all galaxies with signs of morphological perturbations
and interactions. Finally, galaxies that cannot be securely placed
in these morphological classes are labelled as unclassifiable. 

In the text below, whenever we consider only two broad morphological classes we collectively refer to classes 1 and 2 (i.e.
bulges or bulges + faint disk) as `early-type galaxies' (ETGs) and
classes 3 and 4 (i.e. regular and irregular disks) as `late-type
galaxies' (LTGs). Whenever an irregular class is considered, mergers and irregular disks are put together. Some examples of galaxies in these morphological
classes are shown in figure~\ref{fig:visual_stamps}.

\begin{comment}
We note that the visual classification is fairly conservative,
i.e. each time an object inspired any doubt it was put in the
unknown class in order to built a sample as secure as possible to
train the automated algorithms explored in subsequent sections.
\end{comment}

\begin{figure*}
\includegraphics[width=0.99\textwidth]{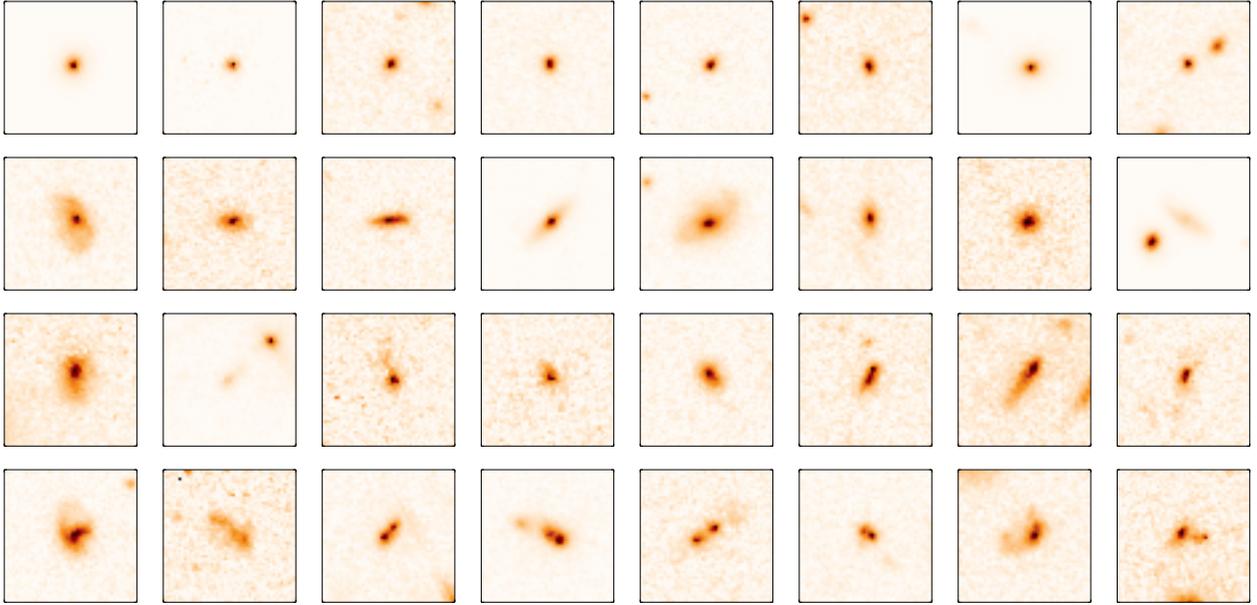}
\caption{H band cutouts of visually classified galaxies as bulges (top row), disks (second row), irregular disks (third row) and mergers (bottom row). The stamp size is $4.5^{"}\times4.5^{"}$. }
\label{fig:visual_stamps}
\end{figure*}

%.........................................................................................................

\subsection{Low-redshift comparison sample}
\label{sec:SDSS} One of the purposes of this study is to explore
the structural evolution of galaxies across cosmic time. To study
this evolution, we will compare the ERS population to a sample of
$\sim14000$ nearby ($z<0.1$) galaxies, drawn from the SDSS, which
have been visually inspected by \citet[][NA10 hereafter]{Nair2010}.
To the NA10 sample, we add a subsample of visually
classified mergers from the Galaxy Zoo project. To
create a set of comparable images, we place the $g$-band images of
the NA10 galaxies at high redshift (following the same redshift and
magnitude distribution of the ERS sample in \citealp{Windhorst2011}) and convolve them with a gaussian to match the ERS FWHM before adding an average ERS
background (see Huertas-Company et al. 2008, 2009 for more
details). Morphological parameters are then estimated from these
simulated images in an identical fashion to the real ERS galaxies
(see section~\ref{sec:morpho_params}). The stamps from the NA10
sample used for this work are available at \url{gepicom04.obspm.fr/}.
We refer readers to \cite{2013MNRAS.435.3444P} for a detailed
description of how the stamps were created. We note that no
physical evolution is included when redshifting these
galaxies, but the observing conditions are preserved,
making this sample well-suited to study the evolution of
morphological parameters from $z\simeq0$ to $z\simeq3$.

\section{Non-parametric morphologies of galaxies at $1<z<3$}
\label{sec:props}

\subsection{Distribution of morphological classes in classical 2-D planes at $1<z<3$}
We begin by exploring the general behaviour of the morphological
parameters for different visually-identified morphological classes
at $z>1$. Figures~\ref{fig:MHC_zle2} and~\ref{fig:MHC_zge2}
present the distribution of visual morphological classes in the
$(C-A)$, $(G-M_{20})$, $(C-S)$, $(C-e)$, $(n-e)$ and $(n-C)$ parameter planes
in two redshift bins ($1<z<2$ and $2<z<3$) respectively.

As is the case at low redshift (e.g Huertas-Company et al. 2011),
the $(C-A)$ plane indicates that spheroids are more
concentrated than the other morphological classes
, but with large scatter ($\tilde{C}_{bulge}\sim2.7\pm0.3$, $\tilde{C}_{disk}\sim2.3\pm0.3$). A simple selection based on concentration (i.e. $C>2.5$) for bulges will hence result in a poor classification since only $70\%$ of the bulges have concentration larger than 2.5 and almost $30\%$ fall also in this region. While the
median asymmetry in the merger class (66\% of mergers have $A>0.2$) is also higher than that in regular
galaxies ($\tilde A_{disks+bulges}\sim0.11\pm0.05$), it is apparent
that many ($>30\%$) late-type galaxies (especially irregular disks) also
exhibit high asymmetry values (see $C-A$ plane). This makes extremely
difficult to cleanly separate these morphological types using
parameters alone (see section~\ref{sec:auto_morph}).

In the $G-M_{20}$ plane, spheroids are again reasonably
well separated in $G$ ($\tilde{G}_{bulge}\sim0.56\pm0.05$ vs.
$\tilde{G}_{disk}\sim0.46\pm0.05$). However, M$_{20}$ does not
appear to provide an efficient route for identifying
irregular/merger objects at $z>1$ at least when only the median
values are considered
($\tilde{M}_{20-regular}\simeq\tilde{M}_{20-irregular}\sim-1.49$ ). A
more detailed analysis is presented in
section~\ref{sec:auto_morph}. The clumpiness parameter $S$ appears
to be slightly more sensitive to irregularities in the galaxy structure as
shown in the $C-S$ plane ($\tilde{S}_{regular}\sim0.02\pm0.02$ and
$\tilde{S}_{regular}\sim0.04\pm0.02$) but not discriminant enough by itself. There are in fact a significant
fraction of \emph{regular} galaxies with relatively high
clumpiness values and also an important number of galaxies for which the clumpiness is close to $0$ (fig.~\ref{fig:MHC_zle2}) of both morphological types. Recall that the clumpiness is close to $0$ when the galaxy has no high frequency structure which means that the light profile is smooth or unresolved.

The ellipticity parameter (see $C-e$ and $n-e$ planes) at $z>1$
behaves similarly to the situation at low redshift, with
disk-dominated galaxies being more elongated than bulges
($\tilde{e}_{bulge}\sim0.18$, $\tilde{e}_{disk}\sim0.36$).
Finally, the Sersic index correlates well with the concentration
as expected and most disk galaxies show Sersic indices lower than
2.5 (similar to their counterparts in the local Universe).
However, it is worth noting that $>30\%$ of spheroids also show such low Sersic values, implying that a
morphological selection based exclusively on this parameter, will
have significant contamination as extensively showed in previous
works (e.g \citealp{2012ApJ...754..141M}).

\begin{figure*}
\includegraphics[width=0.99\textwidth]{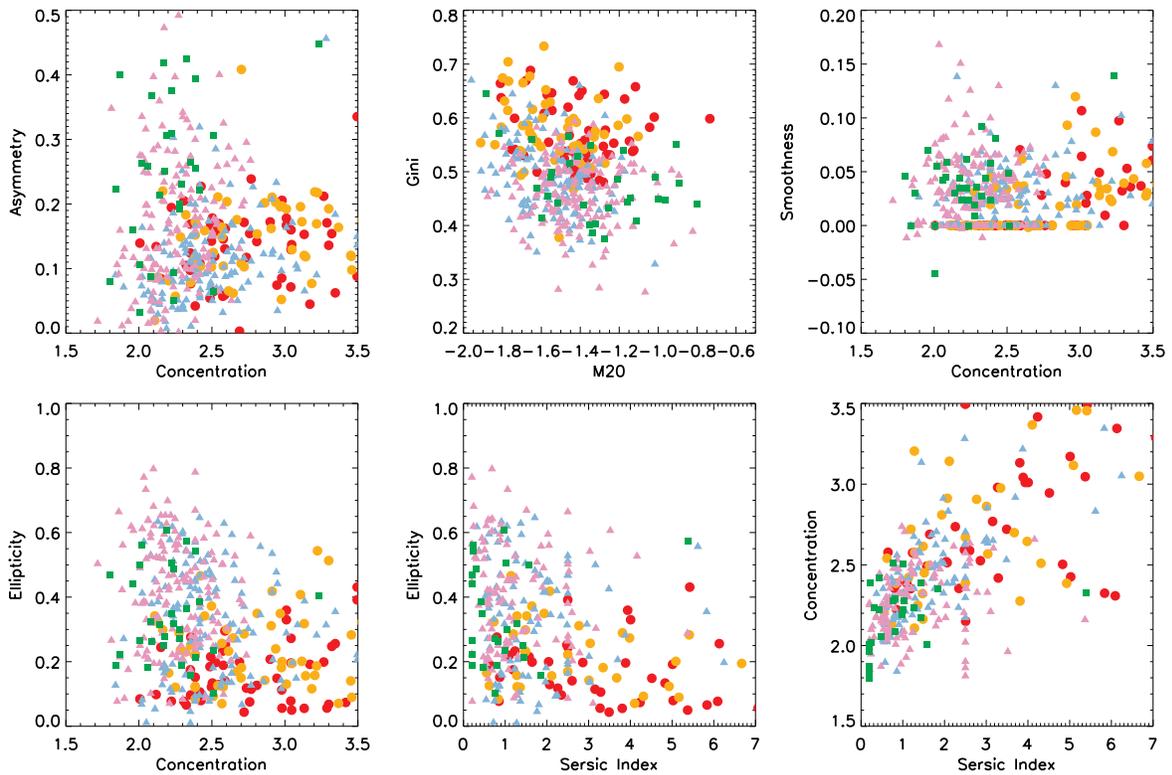}
\caption{2D morphological planes for visually classified galaxies
with $1<z<2$. Red circles are bulges without disk, orange circles
are bulge dominated galaxies with a faint disk, blue triangles are
regular disks, magenta triangles indicate disturbed disks and
green squares show mergers. } \label{fig:MHC_zle2}
\end{figure*}

\begin{figure*}
\includegraphics[width=0.99\textwidth]{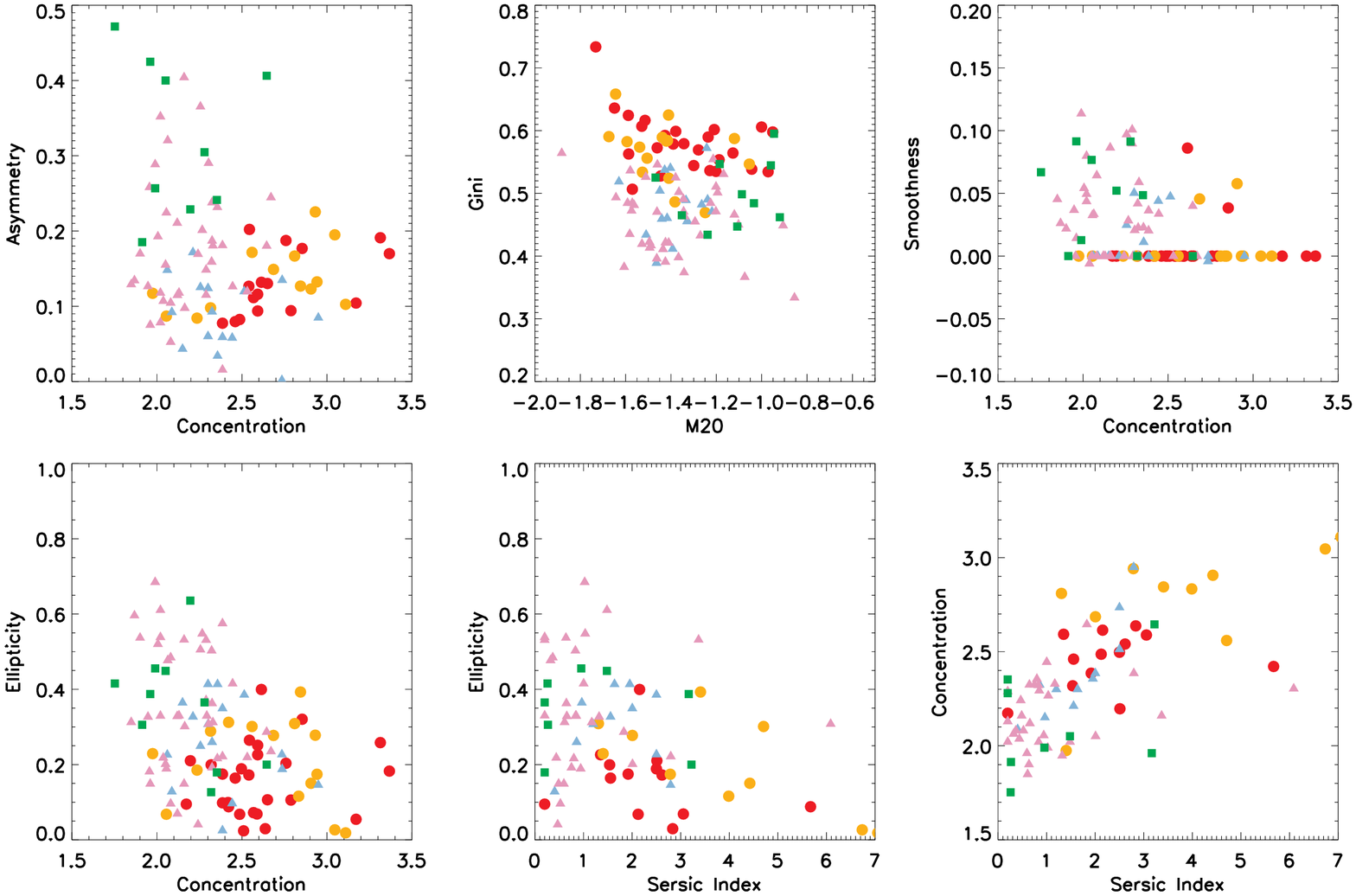}
\caption{2D morphological planes for visually classified galaxies with $2<z<3$. Red circles are bulges without disk, orange circles are bulge dominated galaxies with a faint disk, blue triangles are regular disks, magenta triangles indicate disturbed disks and green squares show mergers.}
\label{fig:MHC_zge2}
\end{figure*}

%.........................................................................................................

\subsection{Dependence of morphological parameters on wavelength}
\label{sec:uv_opt} The high-resolution HST images now available
span both optical and near-infrared filter sets, which trace the
rest-frame UV and optical at $z>1$ respectively at a very similar depth \citep{Windhorst2011} It is desirable
to quantify the statistical variation in the derived values of
morphological parameters with wavelength. In
Figure~\ref{fig:lambda}, we compare values of our morphological
parameters for the same objects, measured in the $F775W$ (i band) filter
(rest-frame UV at $z\simeq2$) and the $F160W$ filter (rest-frame
optical at $z\simeq2$). Since both images do not have the same FWHM ($0.077^{"}$ vs. $0.150^{"}$), we first convolve the $F775W$ with a gaussian filter with a quadrature
difference $FWHM\sim0.125$ to match the resolution of the $F160W$ and get rid of any resolution effect on the derived parameters. We also used the segmentation maps obtained in the H band to compute the morphological parameters in the i band to make sure that we consider the same physical area in each galaxy. Recall that we only show the galaxies for which parameters were properly computed in both photometric bands so the number of objects in each panel might differ.

Overall, morphological parameters exhibit a correlation in the two filters. The parameters that measure concentration ($C$, $G$,
$M_{20}$), exhibit a good correlation with small scatter in the
two filters, with typical variations in the median values of less
than $10\%$ and a Pearson correlation coefficient of
$\rho\simeq0.8$. We notice that \cite{2013ApJ...774...47L} reported a significant systematic offset in the Gini coefficient between rest-frame UV and optical which we do not find here. 
Parameters that are specially sensitive to the internal structure of the galaxy, such as asymmetry ($A$) and
clumpiness ($S$), appear to be more wavelength dependent. Indeed,
the correlation between these parameters measured in the
rest-frame optical and UV appears very weak ($\rho\simeq0.4$), as
indicated in Figure~\ref{fig:lambda}.

In terms of our broad morphological classes (ETGs and
LTGs), we find that later types are more affected by the
morphological k-correction, with LTGs being $\sim50\%$ more
asymmetric and $\sim40\%$ more clumpy in the rest-frame UV than in
the rest-frame optical. We notice however that the median clumpiness also changes very significantly ($>100\%$) between the i and H band for ETGs. This is explained because the median value in te H band is almost $0$ for these objects while there are a few objects in the i band which present larger values which make the relative increase very large. Finally, we also find a trend for irregular and disk galaxies to have lower $M_{20}$ coefficients in the i band as reported by \cite{2013ApJ...774...47L}, which is also a consequence of their more patchy morphology. These results confirm similar results already obtained in the local universe (e.g. \citealp{2007ApJ...659..162T, 2002ApJS..143..113W}) and also highlight the importance of
recalibrating morphological proxies each time a different set of
filters is used. Indeed, it is also worth considering whether a
classification that \emph{combines} the same parameters
measured in different wavelength may yield a more robust automated
classification. We explore this issue in
Section~\ref{sec:multi_lambda}.

It is important to notice an important caveat when interpreting the differences in the morphological parameters from one filter to the other in terms of physical differences, since the parameters are in fact very sensitive to the image quality (S/N, FWHM..., e.g. \citealp{1998ApJ...499..112B,Bershady2000,Lisker2008,Robaina2009,2010ApJ...721...98K}). Even though we have tried to make a fair comparison by matching the resolutions, the pixel scales and by using images of similar depth, galaxies in the i band will appear systematically fainter given the shape of the SED ($i-H$ is always positive for our H band selection) and therefore are observed with a lower S/N.

\begin{figure*}
\begin{center}
$\begin{array}{c c}
\includegraphics[width=0.40\textwidth]{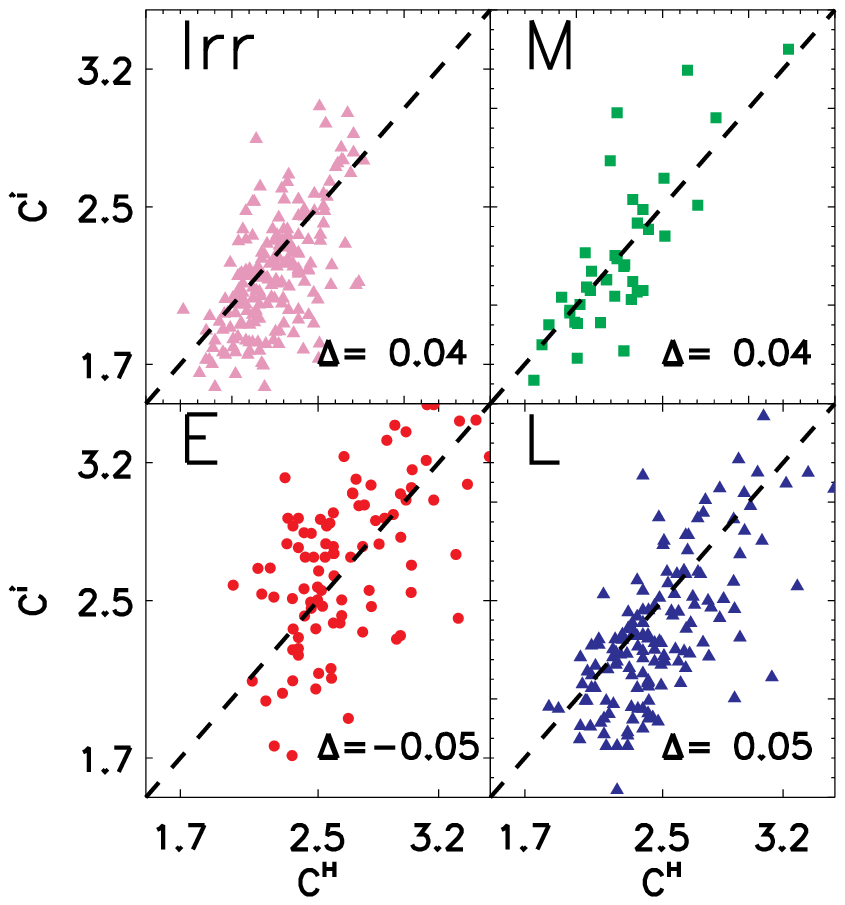} & \includegraphics[width=0.40\textwidth]{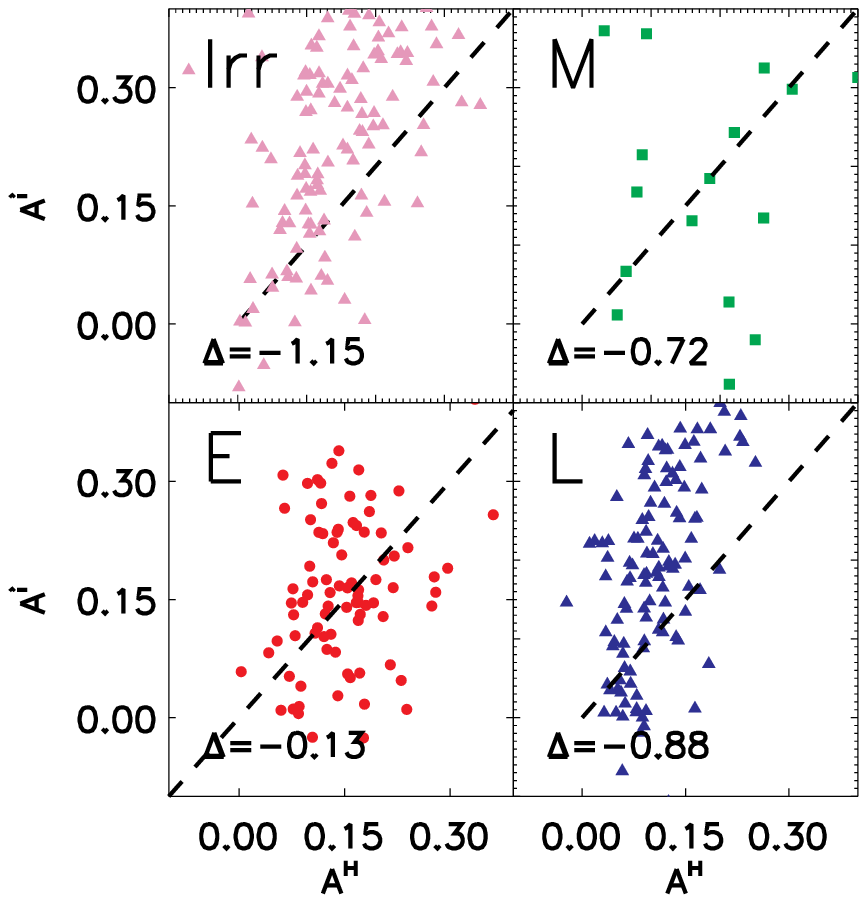}\\
\includegraphics[width=0.40\textwidth]{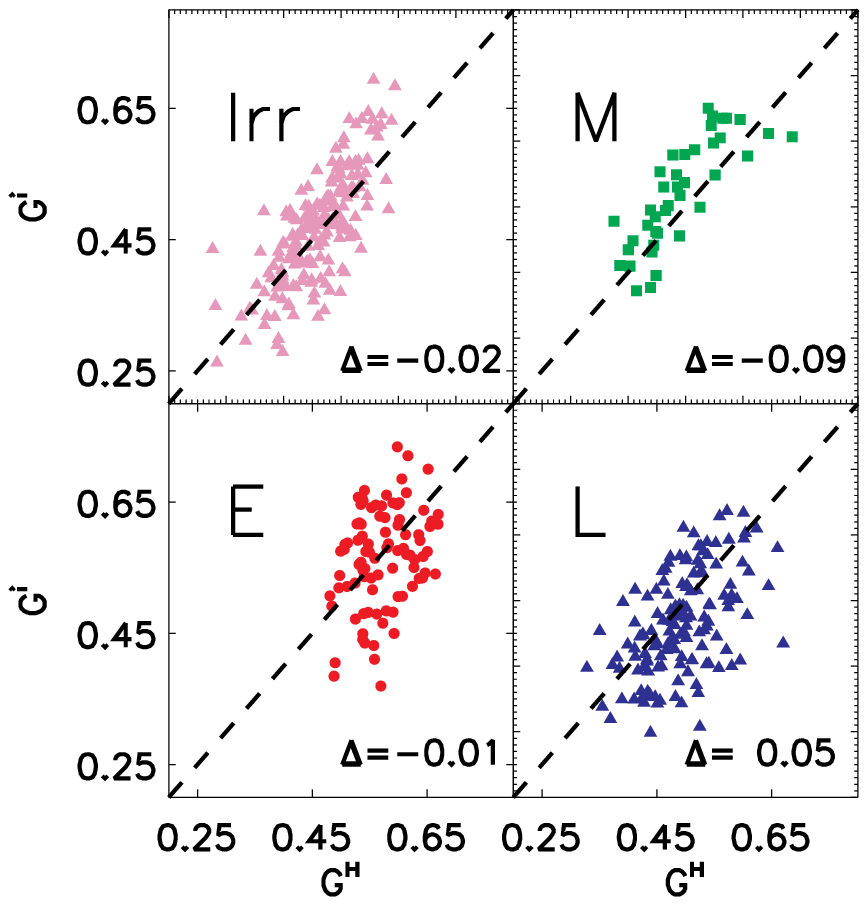} & \includegraphics[width=0.40\textwidth]{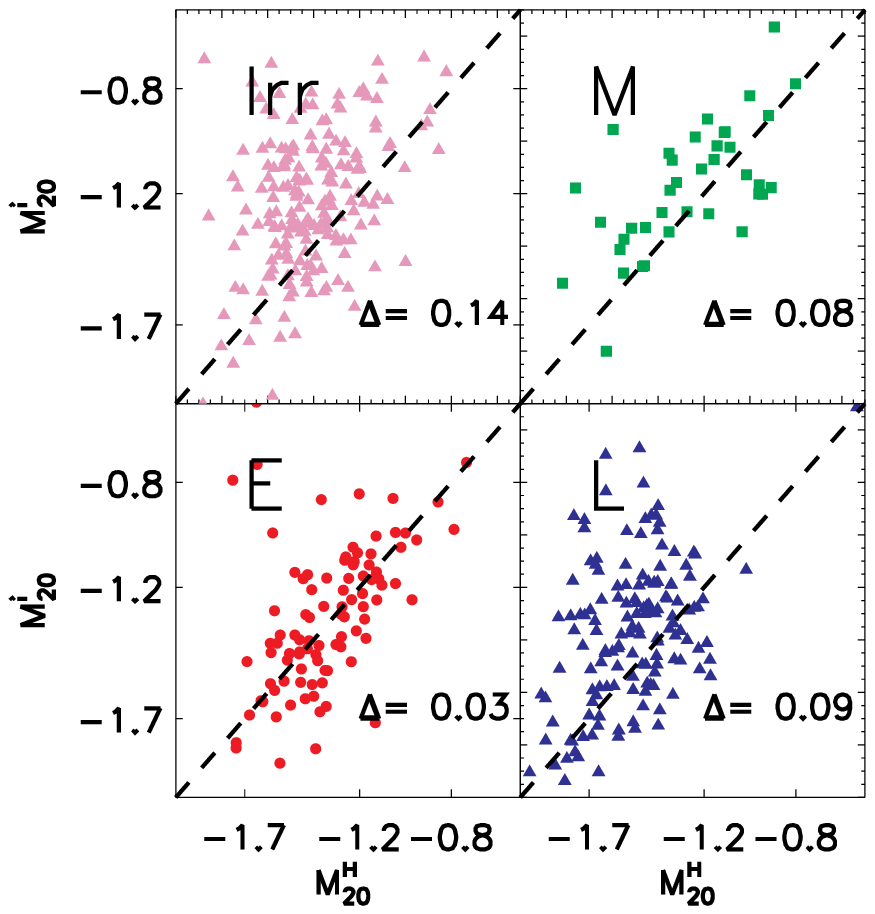}\\
\includegraphics[width=0.40\textwidth]{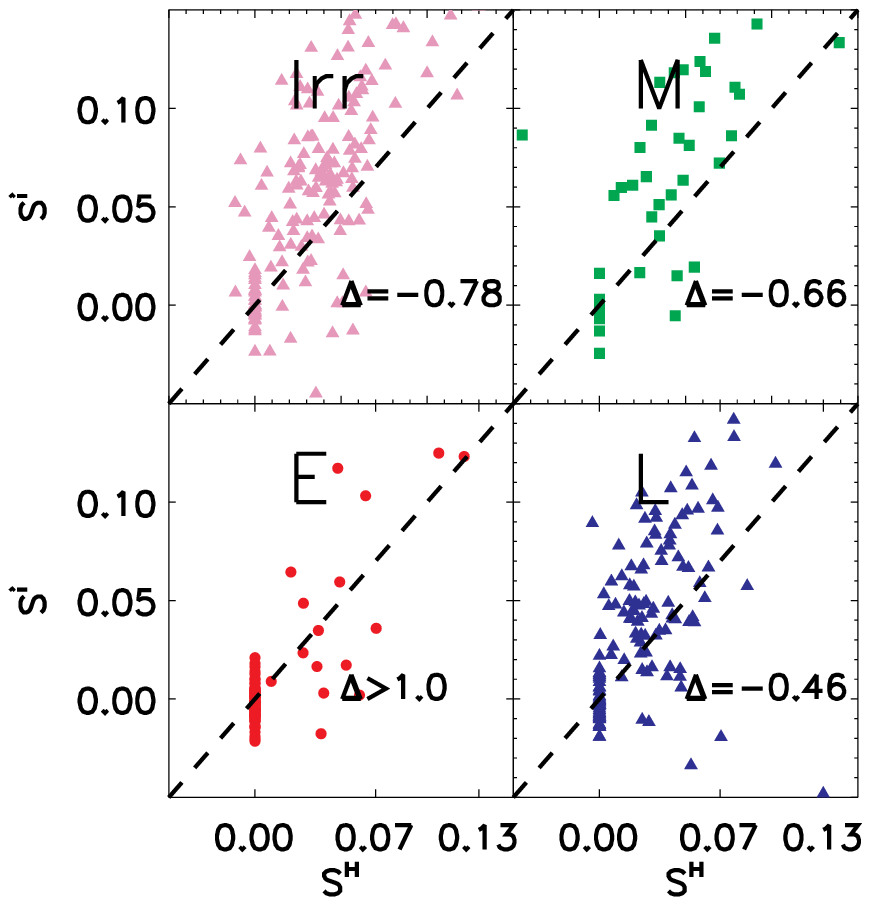} & \includegraphics[width=0.40\textwidth]{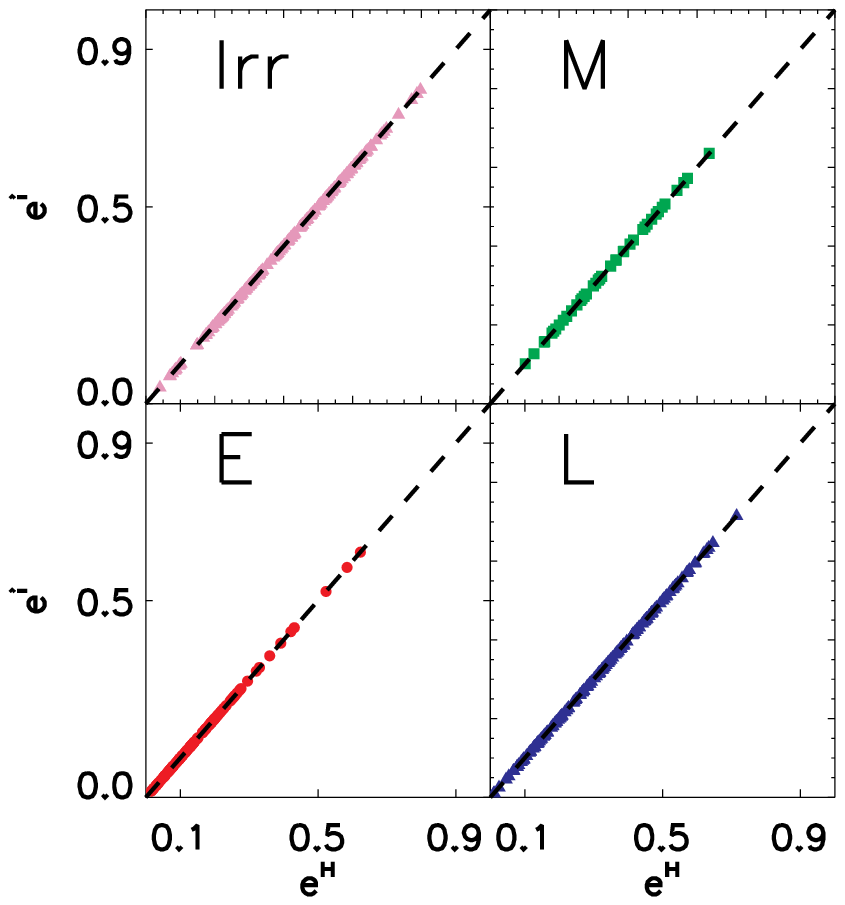}\\

\end{array}$
\caption{Comparison of morphological parameters measured in the i
band and in the H band for different visually identified classes.
Red circles show early-type galaxies, blue triangles regular
late-type galaxies, violet triangles are irregular disks and green
squares indicate mergers. The dashed lines show the one-to-one relation. The values for the ellipticities are identical since we use the same SExtractor ellipse parameters (see text for details). A small offset has been added to the clumpiness (bottom left panel) measured in the i band when it was too close to zero for visibility reasons. We also indicate in each panel the relative difference in the median values between the two filters. } \label{fig:lambda}
\end{center}
\end{figure*}

\begin{comment}
\begin{figure*}
\begin{center}
$\begin{array}{c c}
\includegraphics[width=0.40\textwidth]{C_vs_mag.ps} & \includegraphics[width=0.40\textwidth]{A_vs_mag.ps}\\
\includegraphics[width=0.40\textwidth]{G_vs_mag.ps} & \includegraphics[width=0.40\textwidth]{M20_vs_mag.ps}\\
\includegraphics[width=0.40\textwidth]{S_vs_mag.ps} & \includegraphics[width=0.40\textwidth]{E_vs_mag.ps}\\

%\includegraphics[width=3.3in]{P_bulge.ps}
\end{array}$
\caption{Comparison of morphological parameters measured in the i
band and in the H band for different visually identified classes.
Red circles show early-type galaxies, blue triangles regular
late-type galaxies, violet triangles are irregular disks and green
squares indicate mergers.  } \label{fig:params_mag}
\end{center}
\end{figure*}
\end{comment}

%.........................................................................................................

\subsection{Structural evolution from $z\sim0$ to $z\sim3$}
\label{sec:zevol} In this section, we explore how galaxies in
different morphological classes evolve structurally with look-back
time. In Figure~\ref{fig:highz_local}, we present the same
morphological planes that were explored for the high redshift
sample in the previous section, but this time for the
local sample redshifted to high redshift (as described in
Section~\ref{sec:SDSS}). In this figure, for the high
redshift galaxies, early-types are defined as those classified as
\emph{bulges} or \emph{bulges + faint disks}, late-types are all
\emph{regular disks} and the irregular class contains both
\emph{irregular disks} and \emph{mergers} as defined in
section~\ref{sec:visual}. For the local sample, early-type and
late-type galaxies are galaxies with $T_{type} < 0$ and  $T_{type}
> 0$ according to the \cite{Nair2010} catalog. The irregular
class contains the Galaxy Zoo mergers.

We find that, at first level, the behaviour of the local redshifted sample
is similar to their counterparts at high redshift. Early-types
are typically more concentrated than disk-dominated galaxies and
mergers exhibit higher asymmetry values as would be expected at
any epoch. However, they do not span exactly the same ranges.
Since the measurement of parameters in the local and high redshift
samples are done under the same conditions (i.e. pixel size,
noise, etc.) any differences in the parameter values are
a reflection of the structural changes in galaxies over time.

We quantify this structural evolution in
Figures~\ref{fig:redshift_evolution_E} and
\ref{fig:redshift_evolution_L}. We present the distributions of
values for our morphological parameters, for both the
high-redshift ERS sample and the local redshifted sample split
into early-types and late-types respectively. The relative
differences on the median values are summarized in Table~\ref{tbl:ztable}. Generally,
early-type systems show more evolution in their morphological
parameters than their late-type counterparts. However, parameters measuring the internal structure (A and S) show the largest evolution in the redshift range probed
here ($0<z<3$) regardless of morphological type. Galaxies are
significantly more asymmetric at $z>1$, the increase in asymmetry
being 100\% (a factor of 2) for some early-type systems and LTGs are also significantly more clumpy at $z>1$. The relative variation of the median values of the clumpiness for ETGs shows a puzzling trend at first order. ETGs at $z>1$ are in fact $\simeq100\%$ less clumpy than their local counterparts. This is again explained because a significant fraction of ETGs in the ERS fields have clumpiness values close to 0 (see fig.~\ref{fig:redshift_evolution_E}), which indicate that they have no structures or that they are too compact to resolve them. The
evolution is less pronounced in the remaining parameters, with
variations of $\sim10\%$ for ETGs, and even lower for LTGs.
Early-type galaxies at high redshift appear more
concentrated, have lower $M_{20}$ values and higher Gini
coefficients than their local counterparts. This is simply a
reflection of the fact that early-types are more compact at high
redshift (e.g. \citealp{2005ApJ...626..680D}, Trujillo et al. 2006, Buitrago
et al. 2008, Huertas-Company et al. 2013). Since our
morphological measurements on the high redshift galaxies
are performed under the same conditions as the local
redshifted sample, the evolution in these parameters is a
consequence of real evolution in the early-type population over
cosmic time.

An important implication of our analysis is that a local sample of
galaxies, simply redshifted to $z>1$ might not be a good
template for measuring the morphological properties of galaxies at
these epochs. In Section~\ref{sec:auto_morph} below, we explore
this further, by studying the accuracy of the classifications when
such a local redshifted sample is used to calibrate automated
classifiers (compared to using a real galaxy sample at high
redshift).

\begin{figure*}
\begin{center}
$\begin{array}{c }
\includegraphics[width=0.9\textwidth]{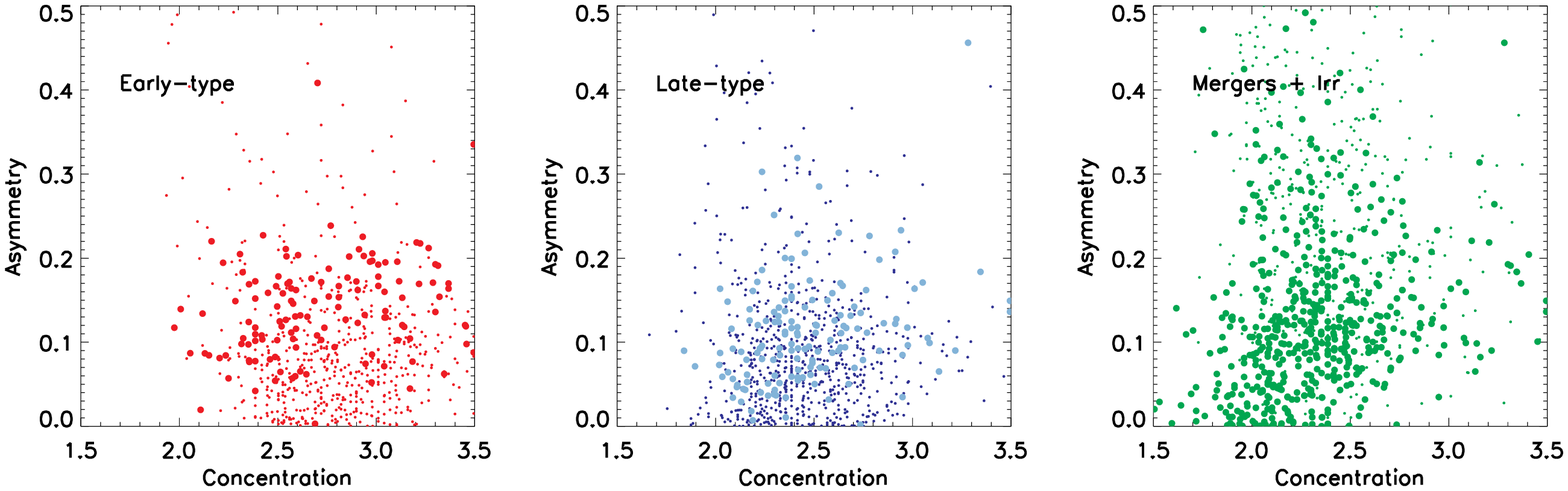} \\
\includegraphics[width=0.90\textwidth]{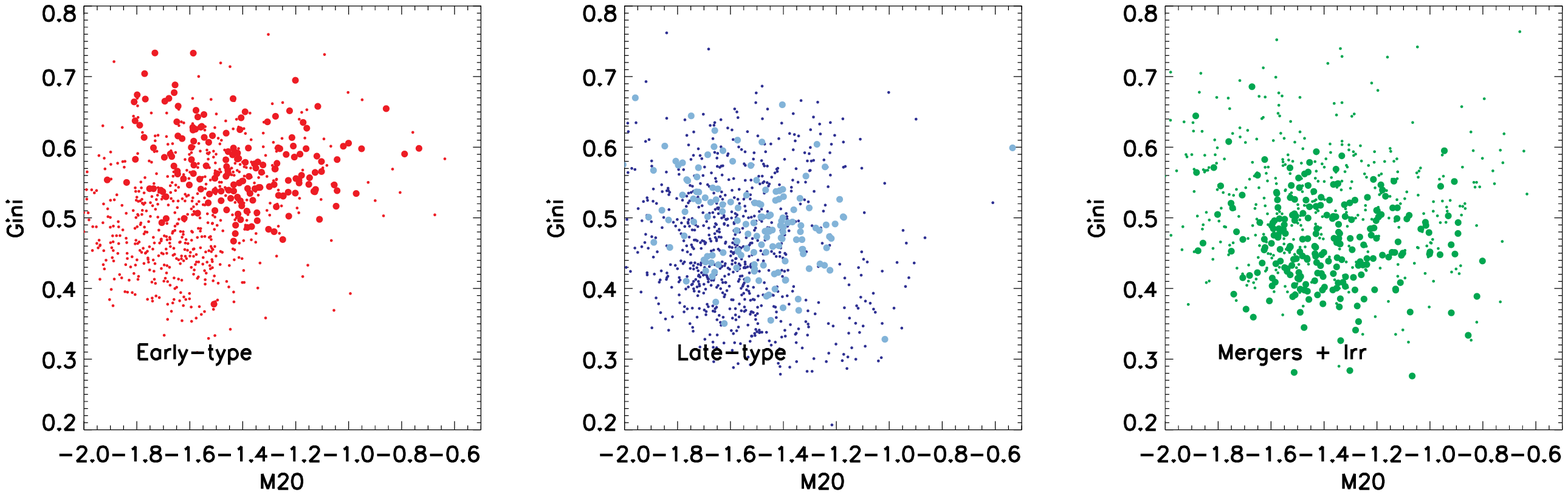} \\
\includegraphics[width=0.90\textwidth]{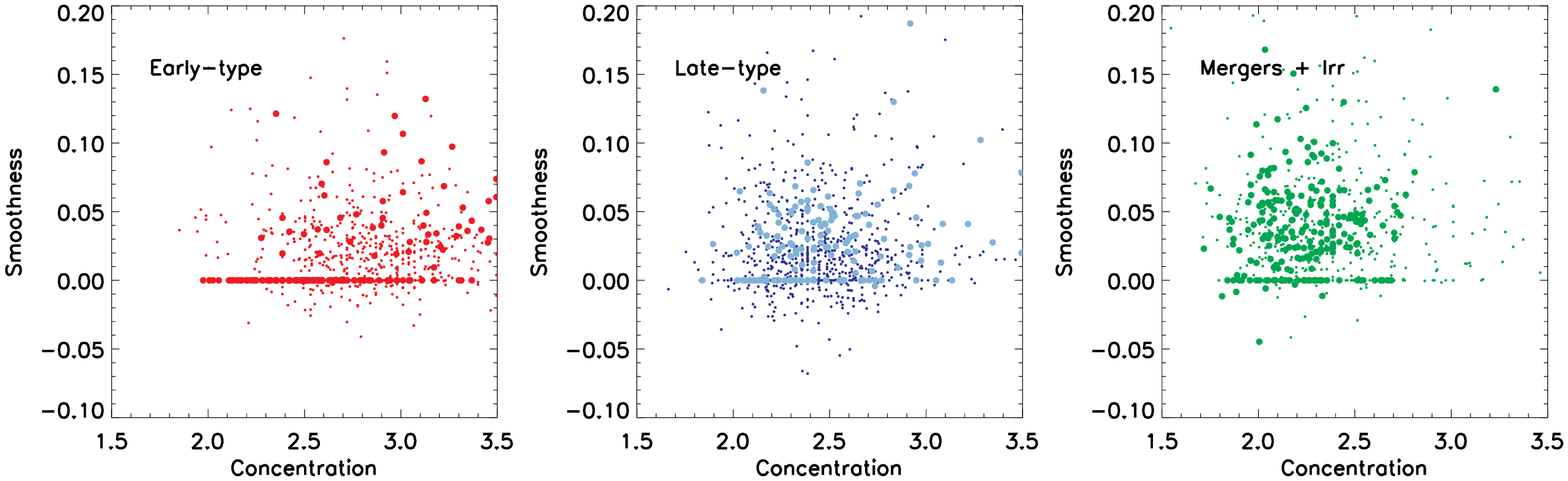} \\
\includegraphics[width=0.90\textwidth]{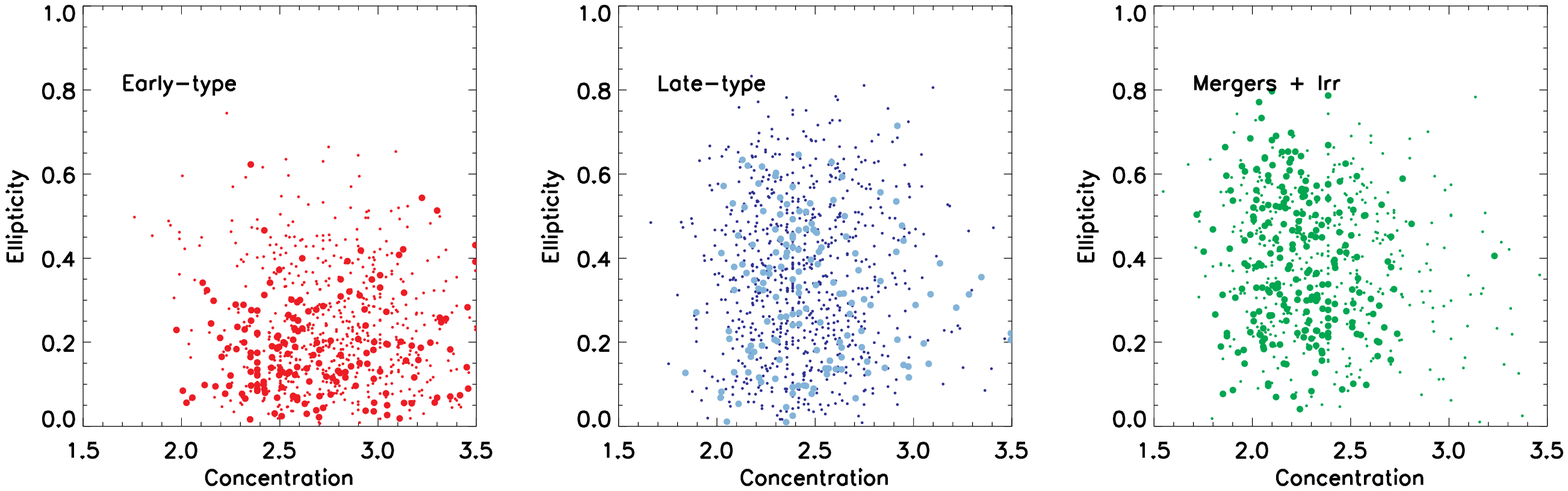} \\

\end{array}$
\caption{Comparison of the distribution in four morphological planes for galaxies at $1<z<3$ (big dots) and local galaxies artificially redshifted (small dots) for three morphological classes as labelled. }
\label{fig:highz_local}
\end{center}
\end{figure*}

\begin{comment}
\begin{figure*}
\includegraphics[width=0.99\textwidth]{plots_morphology_local_redshifted_zle2.ps}
\caption{2D morphological planes for local galaxies from the SDSS redshifted at $1<z<2$. Red circles are early-type galaxies, blue circles are late-type and green triangles are galaxies flagged as mergers.}
\label{fig:morpho_planes_SDSS_1}
\end{figure*}

\begin{figure*}
\includegraphics[width=0.99\textwidth]{plots_morphology_local_redshifted_zge2.ps}
\caption{2D morphological planes for local galaxies from the SDSS redshifted at $2<z<3$.Red circles are early-type galaxies, blue circles are late-type and green triangles are galaxies flagged as mergers.}
\label{fig:morpho_planes_SDSS_2}
\end{figure*}

\end{comment}

\begin{table}
\begin{center}
\begin{tabular}{c c c}

        Param    & ETGs & LTGs  \\
        \hline
        \hline
$\Delta C$ & $0.15\pm 0.04$ & $-0.04\pm 0.02$\\
$\Delta A$ & $ 0.95\pm 0.14 $& $ 0.69\pm 0.08$\\
$\Delta G$ & $ 0.16\pm 0.02 $& $ 0.04\pm 0.01$\\
$\Delta M_{20}$ & $ -0.12\pm 0.02$ & $ -0.08\pm 0.01$\\
$\Delta S$ & $ -1.00\pm 0.13$ & $1.01\pm 0.18$\\
$\Delta e$ & $ -0.10\pm 0.07$ & $ 0.01\pm 0.06$\\

\end{tabular}
\caption{Relative variation of different morphological parameters between $z\sim 0$ and $z\sim3$ early-type (left column) and late-type (right column) galaxies. $(\Delta X=(\tilde{X_{z>1}}-\tilde{X_{z=0}})/\tilde{X_{z=0}})$.}
\label{tbl:ztable}
\end{center}
\end{table}

\begin{figure*}
\begin{minipage}{172mm}
\begin{center}
\includegraphics[width=0.85\textwidth]{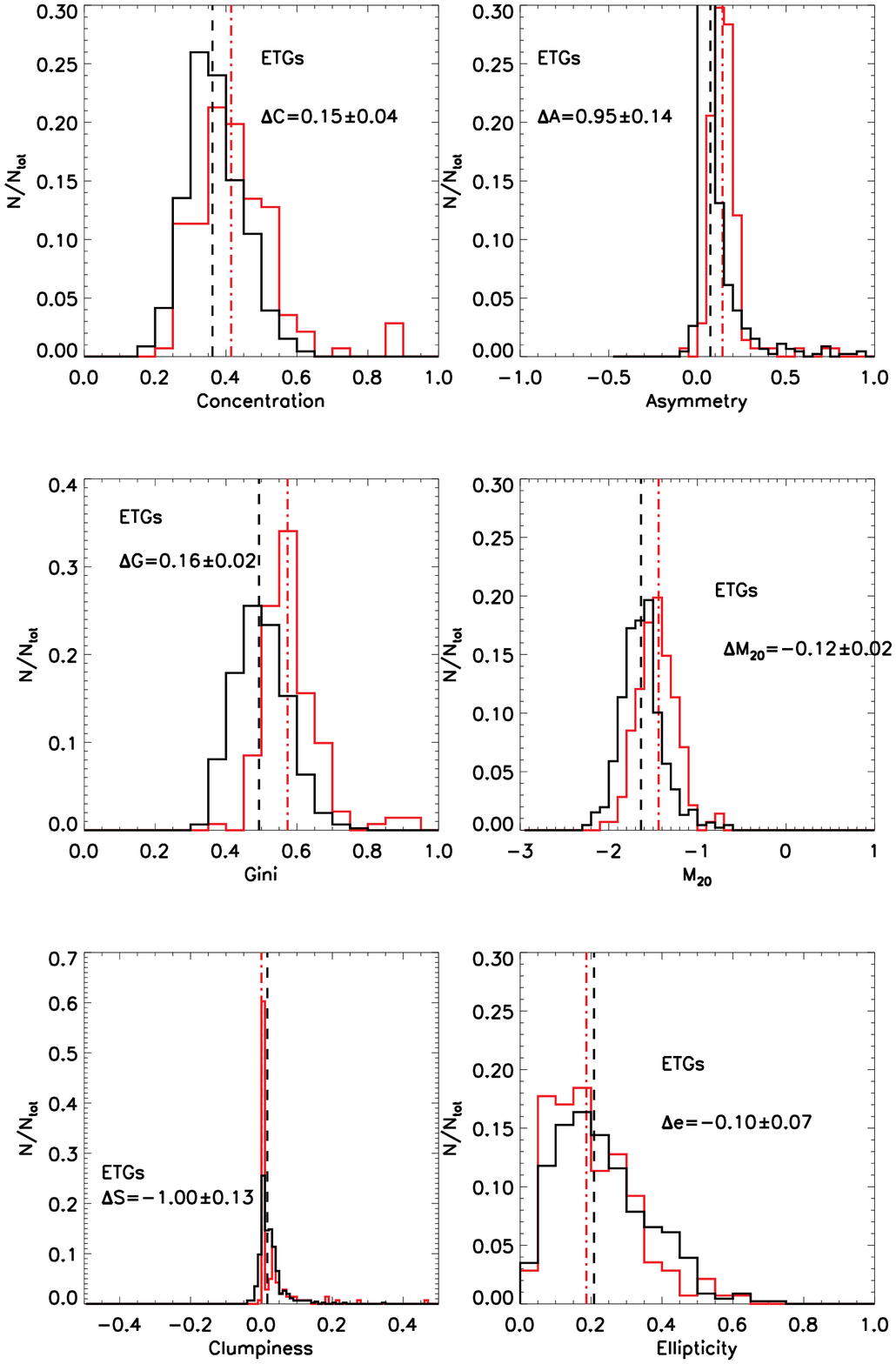}
\caption{Distribution of morphological parameters for high redshift ETGs (red line) and local galaxies (black line) redshifted at the same redshift as explained in the text. For each parameter we provide the relative variation of the median value between $z\sim3$ and $z\sim0$, i.e. $\Delta = \frac{median_{z>1}-median_{z=0}}{median_{z=0}}$} \label{fig:redshift_evolution_E}
\end{center}
\end{minipage}
\end{figure*}

\begin{figure*}
\begin{minipage}{172mm}
\begin{center}
\includegraphics[width=0.85\textwidth]{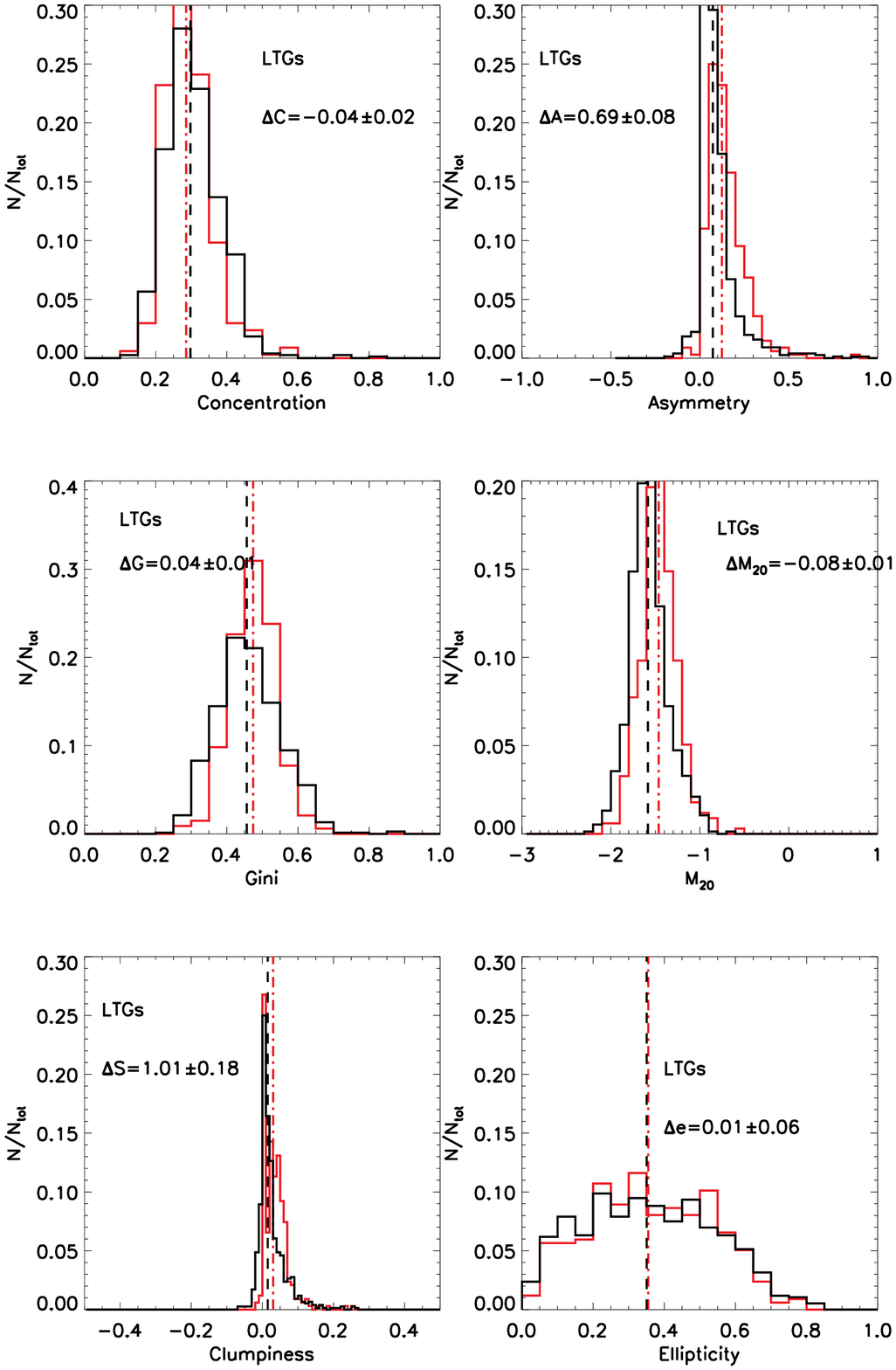}
\caption{Distribution of morphological parameters for high redshift LTGs (red line) and local galaxies (black line) redshifted at the same redshift as explained in the text. For each parameter we provide the relative variation of the median value between $z\sim3$ and $z\sim0$, i.e. $\Delta = \frac{median_{z>1}-median_{z=0}}{median_{z=0}}$} \label{fig:redshift_evolution_L}
\end{center}
\end{minipage}
\end{figure*}

\begin{comment}
\begin{figure*}
\includegraphics[width=0.99\textwidth]{E_stamps_lambda.ps}
\caption{H and i band cutouts of the same galaxies classified as regular disks visually by MHC and SK.}
\label{fig:visual_stamps_E_lambda}
\end{figure*}

\begin{figure*}
\includegraphics[width=0.99\textwidth]{L_stamps_lambda.ps}
\caption{H and i band cutouts of the same galaxies classified as bulges visually by MHC and SK.}
\label{fig:visual_stamps_L_lambda}
\end{figure*}

\begin{figure*}
\includegraphics[width=0.99\textwidth]{Lirr_stamps_lambda.ps}
\caption{H and i band cutouts of the same galaxies classified as irregular disks visually by MHC and SK.}
\label{fig:visual_stamps_Lirr_lambda}
\end{figure*}

\begin{figure*}
\includegraphics[width=0.99\textwidth]{merger_stamps_lambda.ps}
\caption{H and i band cutouts of the same galaxies classified as mergers visually by MHC and SK.}
\label{fig:visual_stamps_M_lambda}
\end{figure*}
\end{comment}

%.........................................................................................................

%\begin{table}
%\begin{center}
%\begin{tabular}{l|c}

   %      & Completeness & Contaminations \\\hline
   % ETGs & 77 (28) & $C_{LTGs}=30, C_{mergers}=2, C_{DETGs}=18$\\
   % ETGs with disk & 0.21 (18)\\
   % LTGs & 0.27 (120)\\
   % mergers & 0.82 (24)\\
     % \end{tabular}
%\caption{Completeness and purities of the different morphological types}
%\end{center}
%\end{table}

%.........................................................................................................

\section{Automated bayesian morphology at $z>1$ with galSVM}
\label{sec:auto_morph}
 In this section we quantify the reliability
of automated morphological classifications at $z>1$.
Traditionally, combinations of morphological parameters have been
shown to be sensitive to different morphological classes. For
example, the $G$-M$_{20}$ plane has been used to identify merger
candidates at low and intermediate redshift \cite[see
e.g.][]{Lotz2004}, while the $C$-$A$ plane has been used to
discriminate between early-type and late-type galaxies
\citep[e.g.][]{Conselice2003, Huertas-Company2008}. Since the reliability of
individual parameters can depend somewhat on factors such as
signal-to-noise ratio, brightness and galaxy size
\citep[e.g.][]{1998ApJ...499..112B,Bershady2000,Lisker2008,Robaina2009,2010ApJ...721...98K},
a multi-dimensional analysis that combines several parameters
maximizes the accuracy of the parameter-based classification
\citep{Huertas-Company2008,Huertas-Company2011}.

The galSVM\footnote{\url{http://gepicom04.obspm.fr/galSVM/Home.html}} code
uses the information contained in several morphological parameters
simultaneously, to associate a morphological class to an
individual galaxy, by means of a Support Vector Machine (SVM). For
this particular work, the galSVM classification simultaneously
uses 7 parameters ($C$, $G$, $M_{20}$, $A$, $S$, $n$ and $e$). All
parameters are measured in the $F160W$ filter (but in
section~\ref{sec:multi_lambda} we explore the effects of adding
parameters measured in shorter-wavelength bands).

Since the technique is optimized to classify only 2 morphological classes at a time, we perform two separate
classifications. The first is aimed at distinguishing between
broad morphological classes (LTG vs. ETG). The second attempts to
differentiate irregular and merging galaxies from the regular
classes. In essence, each galaxy in the sample is
allocated two probabilities: $P_{ETG}$ (probability of being an
early-type galaxy) and $P_{irr}$ (probability of being an
irregular disk). galSVM is calibrated using a training sample of
visually-classified morphologies, with the assumption that the
morphologies in the training set are robust. For an individual
galaxy, the code then measures an a posteriori
probability for that object to belong to a given morphological
class.

We note that a key requirement for accurate classification is that
the training sample should be as close as possible to the real
sample one wishes to classify. An approach used in previous works
(e.g Huertas-Company et al. 2008, 2009) has been to use a local
sample with high quality morphologies which are redshifted to the
epoch of interest. However, our results in Section 5.2 indicate
that, due to the evolution of morphological parameters between
$z>1$ and the local Universe, redshifted local samples do not
provide a good training set for studying galaxy morphologies in
the early Universe.

For that reason, in this work we use 2 different training sets to
train galSVM in order to calibrate the effects of the training on
the final classification at $z>1$:\\

\noindent \textbf{SDSS sample:} Our first training set is the
local SDSS comparison sample described in Section 2.2. Recall
that, to create this comparison sample, SDSS galaxies from NA10 are
redshifted, scaled in luminosity to account for cosmological
dimming and placed in a real ERS background with the proper pixel
scale. As noted before, the drawback of using this training set is
that no morphological evolution is assumed i.e. the morphological
structure of the redshifted dataset are identical to the local
galaxy population. This approach has been used at $z<1$ without
significant biases (e.g. \citealp{2013MNRAS.435.3444P, 2013MNRAS.428.1715H}). However, as shown in section~\ref{sec:zevol}, there is
significant evolution of the morphological parameters between
$z\sim0$ and $z\sim3$, which might induce important effects in the
final classification at $z>1$. By employing this training set, we
wish to explicitly calibrate the effects of using redshifted local
samples as training sets for morphological work in the early
Universe.\\

\noindent \textbf{High redshift sample:} As described in
section~\ref{sec:visual}, the main sample used for this work has
been visually classified directly on the H-band ERS
images. This training set should be considered the
\emph{best case}, since we are using data at the same epoch for
both the training and classification. Note, however that there is
a certain amount of redundancy in this exercise, since there is a significant overlap between the datasets used for both the training and the automated
classification. To explore the impact of this redundancy, we check
our classification results against the visual classification
recently performed by the CANDELS team
(\citealp{2014arXiv1401.2455K} - see section~\ref{sec:candels}).

\subsection{Classification of broad morphological types}
We begin by investigating how broad morphological classes, i.e.
bulge-dominated and disk-dominated galaxies, are recovered using
galSVM. For galaxies visually classified as LTGs and ETGs,
Figure~\ref{fig:bayesian_probs_E_L} shows the probability of these
galaxies to be classified as early-type by galSVM. The left-hand
panel shows results obtained using a high-redshift galaxy sample
as the training set, while the right-hand panel shows the
corresponding distributions with the local redshifted sample used
as a training set.

For the results obtained using the high-redshift training set, we
find good agreement between visual and automated classifications
at $z>1$. The probability distribution for visually-classified
ETGs clearly peaks at high values of p$_{ETG}$ from galSVM, while
visually-classified LTGs show a peak at low values of p$_{ETG}$.
When the SDSS sample is used for training (right panel), we still
see the two peaks but the distributions exhibit more
extended wings. This is especially true for the early-type
population, which contains a considerable fraction of galaxies
with low p$_{ETG}$ probabilities.

As noted before, the advantage of an SVM-based classification is
that it provides a \emph{probability} rather than a binary
classification. It is therefore possible to tune the selection of
a particular morphological class using different probability
thresholds, depending on the actual science goals. The most
natural selection would be to include in a given class, all
objects that have a probability greater than 0.5, but it is
possible to change the threshold depending on the properties (e.g.
purity/completeness) required of the sample.

We proceed by quantifying the `purity' and `completeness' of the
galSVM classifications,taking the visual classifications as
reference. For a given probability threshold ($p_{th}$), we define the following quantities:

\begin{itemize}
\item {\it True positives (tp):} \# galaxies with $p_{ETG}>p_{th}$ ($p_{LTG}>p_{th}$)  which are visually classified as early-type (late-type).
\item {\it True negatives (tn):} \# galaxies with $p_{ETG}<p_{th}$ ($p_{LTG}<p_{th}$) which are visually classified as late-type (early-type).
\item {\it False positives (fp):} \# galaxies with $p_{ETG}>p_{th}$ ($p_{LTG}>p_{th}$) which are visually classified as late-type (early-type).
\item {\it False negatives (fn):}  \# galaxies with $p_{ETG}<p_{th}$ ($p_{LTG}<p_{th}$) which are visually classified early-type (late-type).
\end{itemize}

The purity P, also known as reliability (the fraction of well-classified objects among all
objects classified in a given class) and the completeness C (the
fraction of well-classified objects among all objects really
belonging to a given class) are then defined as follows:

\begin{equation}
P=1-\frac{fp}{fp+tp}
\end{equation}

\begin{equation}
C=\frac{tp}{fn+tp}
\end{equation}

P (Purity) effectively measures the level of contamination in the
automated classification. For example, if $30\%$ of the galaxies
classified by galSVM as ETGs are, in fact, classified as LTG via
visual inspection, then the purity of the galSVM-classified sample
will be $70\%$. C (Completeness), on the other hand, measures how
well galSVM recovers objects in a visually-classified class. For
example, if $C=100\%$ for a sample of galaxies classified as ETGs
by galSVM, then this indicates that all galaxies visually
classified as ETGs are recovered as such by the automatic
classification. A perfect classification is, therefore, $100\%$
pure and $100\%$ complete. In practice, there is a tradeoff
between the two quantities, since requiring better completeness
typically involves increasing the level of contamination in the
automated results.

In Table~\ref{tbl:ETG_C_P} we present the completeness and purity
of samples selected with different probability thresholds, using
the local and the high-redshift training sets. Not unexpectedly,
an increase in the probability threshold results in a purer sample
while the completeness decreases. This is particularly true for
the ETG population. To achieve a purity greater than $80\%$, we
need to select galaxies with $p_{ETG}>0.7$, for which the completeness
then drops to $\sim60\%$. Generally speaking, late-type galaxies
are more easily recovered with automated proxies. For example,
selecting galaxies with $p_{LTG}>0.4$ leads to a sample that is
90\% complete and $90\%$ pure. This is largely a consequence of
spatial resolution. Galaxies that are clearly extended are easily
identified as disk dominated galaxies. However, the most compact
disk galaxies tend to contaminate the early-type population. This
effect might be reduced if higher resolution images are used. The
ERS images have been resampled to a pixel size of $0.09^{"}$ but
images resampled at a resolution of $0.06^{"}$ are now available
which might help in resolving compact LTGs. This will be explored
in a future work. It is worth noting though that, given that the
number of late-type systems at $z>1$ is significantly higher than
the number of early-type galaxies, the misclassification of a
small proportion of LTGs has a relatively insignificant impact on
the recovery of the general population of LTGs. Some examples of
the galaxies classified automatically are shown in
figure~\ref{fig:auto_stamps}.

In Table~\ref{tbl:ETG_C_P}, we quantify the effect of the training
set used for the galSVM classifications. The accuracy of the
automated classification decreases when a redshifted sample from
the SDSS is used for training, as opposed to a sample of real
high-redshift galaxies. This is expected since, as described in
Section~\ref{sec:zevol}, the morphological parameters used for the
classification show some evolution from low to high redshift.
\emph{Thus, local galaxies are not the optimal template for
classifying a high-redshift ($z>1$) sample.} This discrepancy
between a local and (real) high-redshift sample is particularly
critical for the SVM, which is a machine learning technique that
is particularly sensitive to the training set.
Interestingly, the effect of the training set is reflected
more in the completeness - for example, a threshold value
of $p>0.6$ in the ETG probability yields a completeness of $57\%$
when measured using a low-redshift SDSS training set, while
$\sim70-80\%$ is achieved using a training set of actual
high-redshift galaxies.

Recall that an increase in the probability threshold results in a
purer sample which is typically less complete. A low completeness
might have critical consequences for the scientific conclusions
derived, if the selected sample is biased towards a specific
population. For instance, one might wonder if galaxies with higher
probability values are the brightest and/or the largest and/or the
ones with higher surface brightnesses, as a result skewing the
science results. In Figure~\ref{fig:props_probas} we address this
issue by plotting the size (from galfit), magnitude and surface-brightness
distributions of early and late-type samples selected with
increasing probability thresholds (0.3, 0.5 and 0.7). The main
conclusion is that there is no apparent bias in the main
properties of the selected samples i.e the distributions are
similar ($P_{K-S}>0.8$ - see fig.~\ref{fig:props_probas}),
independent of the probability threshold applied. This result
reflects the fact that the morphological parameters used for the
automated classification are reliable enough in the ranges of
magnitudes, surface brightnesses and sizes explored in this study
(note that we are limited to $H<24$, where visual classifications
are reliable), that the probability is not correlated with any of
this physical parameters.

\begin{figure*}
\begin{center}
$\begin{array}{c c}
\includegraphics[width=3.0in]{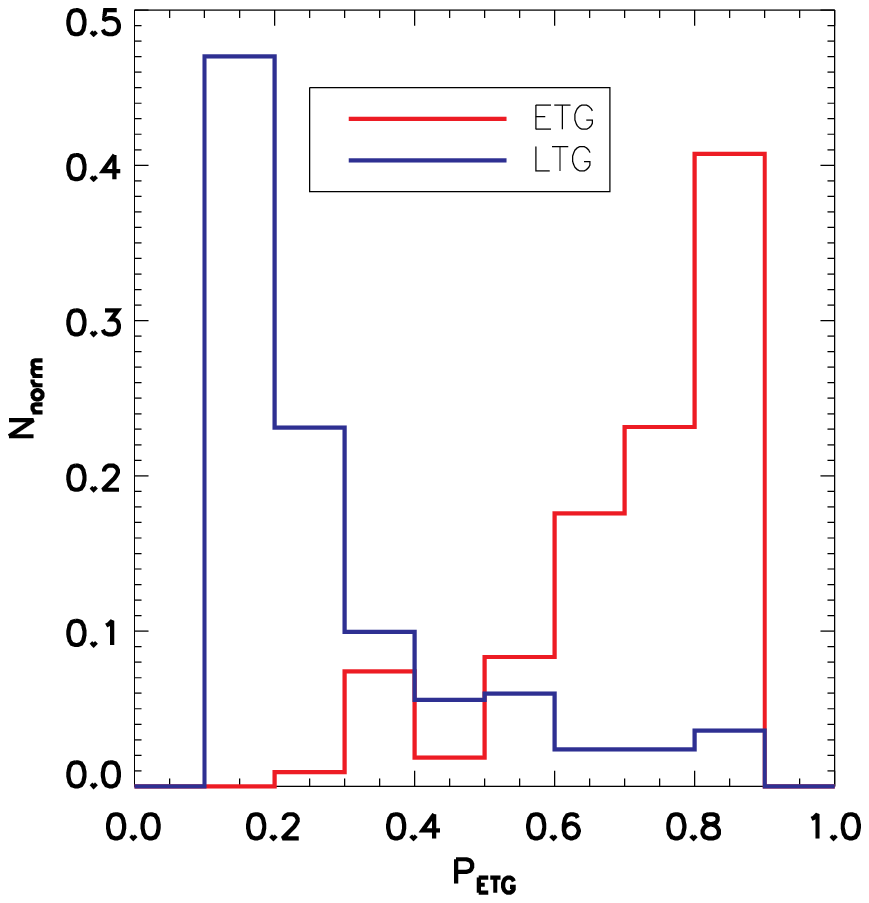} & \includegraphics[width=3.0in]{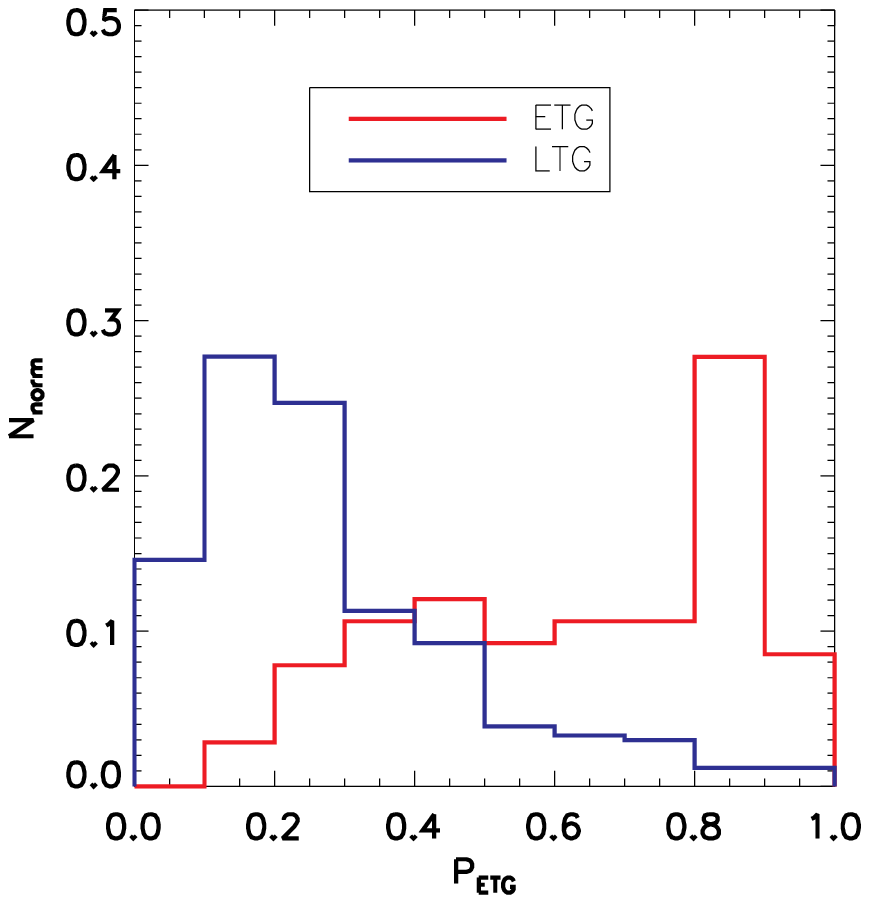}\\

\end{array}$
\caption{ETG Probability distribution for visually classified early-type (red) and late-type (blue) galaxies. The left panel shows the distributions obtained with a high-z training set and the left panel those obtained with a local sample (see text for details).}
\label{fig:bayesian_probs_E_L}
\end{center}
\end{figure*}

\begin{figure*}
\includegraphics[width=0.99\textwidth]{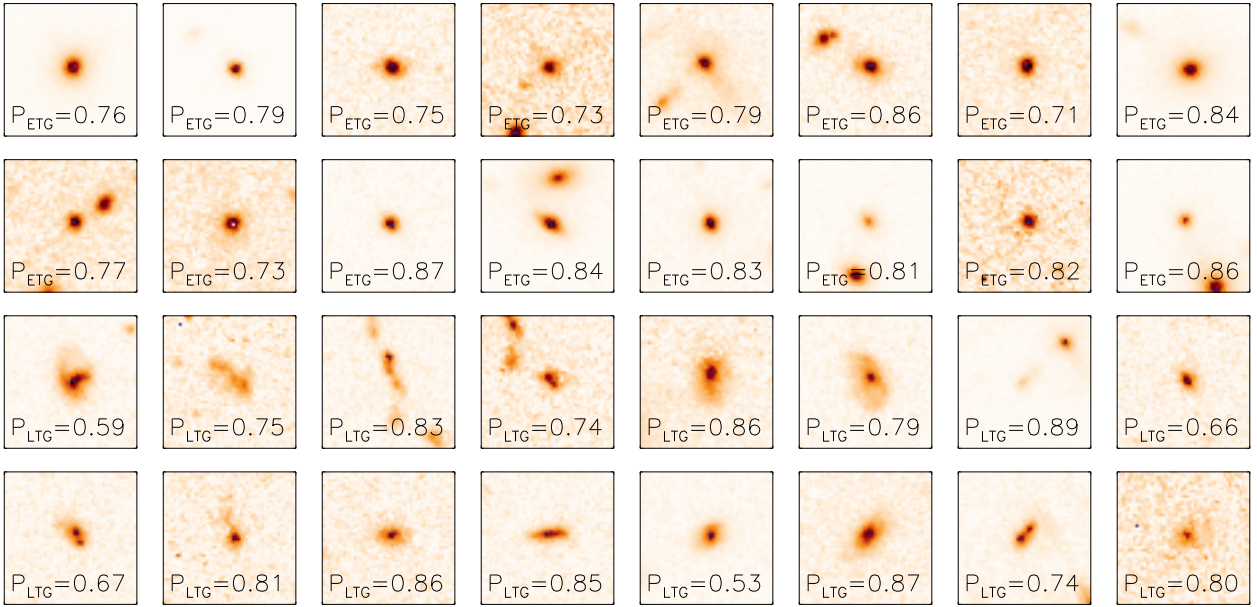}
\caption{H band cutouts of automatically classified early-type galaxies (2 top rows) and late-type (2 bottom rows). ETGs (LTGs) are selected with a probability threshold of $p>0.7$ ($p>0.5$) which results in a contamination of $\sim 15\%$ ($<10\%$). See text for details. The associated probability is indicated in each individual stamp.}
\label{fig:auto_stamps}
\end{figure*}

\begin{comment}
\begin{figure*}
\includegraphics[width=0.99\textwidth]{LTGs_stamps.ps}
\caption{H band cutouts of automatically classified late-type galaxies selected with a probability threshold of $p>0.5$ which results in a contamination of $<10\%$. The associated probability is indicated in each individual stamp..}
\label{fig:vstamps_L_lambda}
\end{figure*}
\end{comment}

\begin{table*}
\begin{center}
\begin{tabular}{c c c c c}

        $P_{thresh}$            & $P^{ERS}(ERS)$ & $C^{ERS}(ERS)$ & $P^{SDSS}(ERS)$ & $C^{SDSS}(ERS)$ \\
        \hline
        \hline
        {\bf ETGs} &&&\\
        \hline
0.3  & 53.68 & 99.03 & 48.80 & 89.71 \\
0.4  & 62.50 & 92.23 & 56.61 & 78.68\\
0.5  & 70.45 & 90.29 & 66.42 & 66.91 \\
0.6  & 78.70 & 82.52 & 71.56 & 57.35 \\
0.7  & 80.00 & 66.02 & 77.11 & 47.06 \\
0.8  & 83.02 & 42.72 & 85.96 & 36.03  \\
\hline
 {\bf LTGs} &&&\\
 \hline
 0.3 & 88.41 & 94.01 & 83.18 & 95.77  \\
0.4  & 93.55 & 91.90 & 85.67 & 92.61  \\
0.5  & 96.08 & 86.27 & 88.07 & 88.38 \\
0.6  & 96.60 & 79.93 & 91.46 & 79.23 \\
0.7  & 99.49 & 69.01 & 95.52 & 67.61 \\
0.8  & 100.00 & 45.77 & 97.58 & 42.61 \\
\hline
 {\bf Irr/mergers} &&&\\
 \hline
0.3 & 67.89 & 97.66 & 54.37 & 87.67 \\
0.4 &73.06 & 93.57 & 63.60 & 73.13 \\
0.5 &76.92 & 87.72 & 75.69 & 60.35 \\
0.6 & 81.18 & 80.70 & 79.85 & 47.14 \\
0.7 & 85.11 & 70.18 & 85.06 & 32.60\\
0.8 & 87.25 & 52.05 & 85.42 & 18.06 \\
\end{tabular}
\caption{Completeness (C) and purity (P) for ETGs, LTGs and irregulars at $1<z<3$ in the ERS obtained with a high redshift (ERS, first 2 columns) and low redshift (SDSS, last 2 columns) training sets (see text for details). }
\label{tbl:ETG_C_P}
\end{center}
\end{table*}

\begin{figure*}
\begin{center}
$\begin{array}{c c}
\includegraphics[width=2.8in]{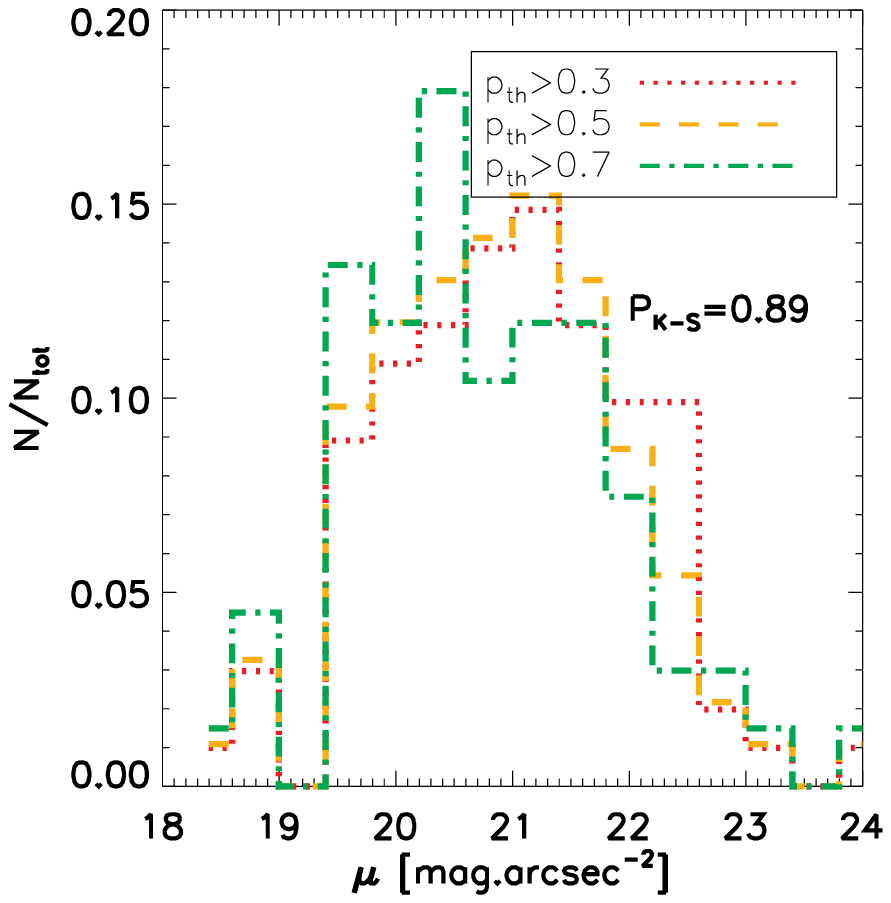} & \includegraphics[width=2.8in]{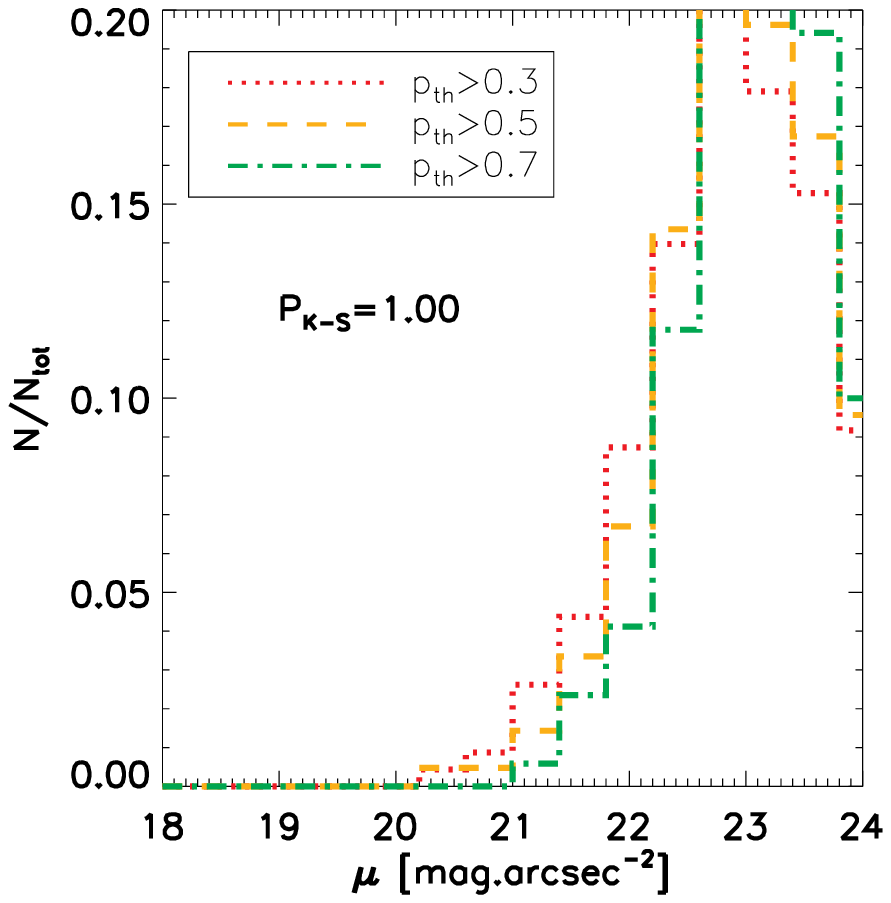}\\
\includegraphics[width=2.8in]{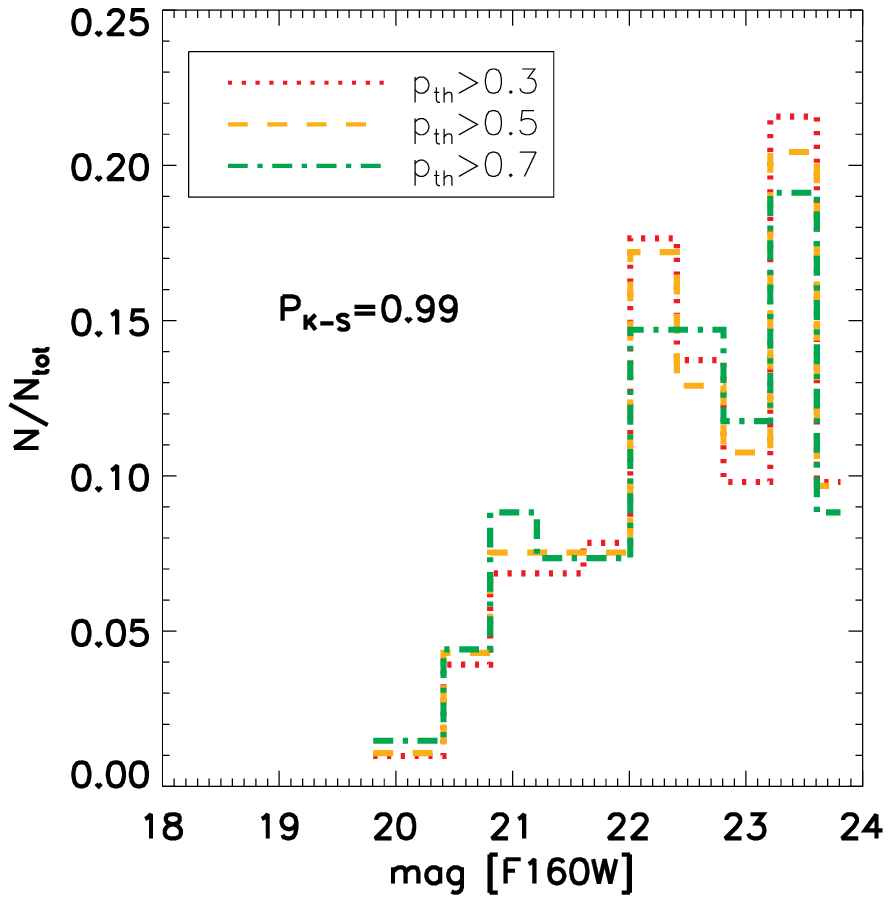} & \includegraphics[width=2.8in]{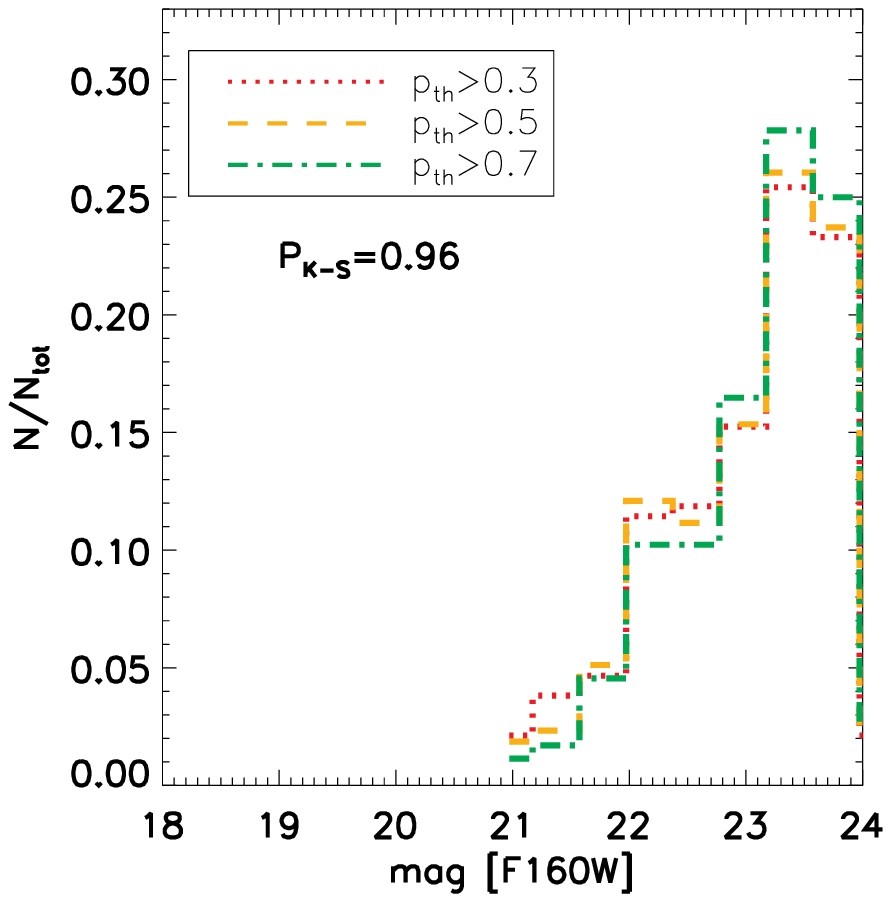}\\
\includegraphics[width=2.8in]{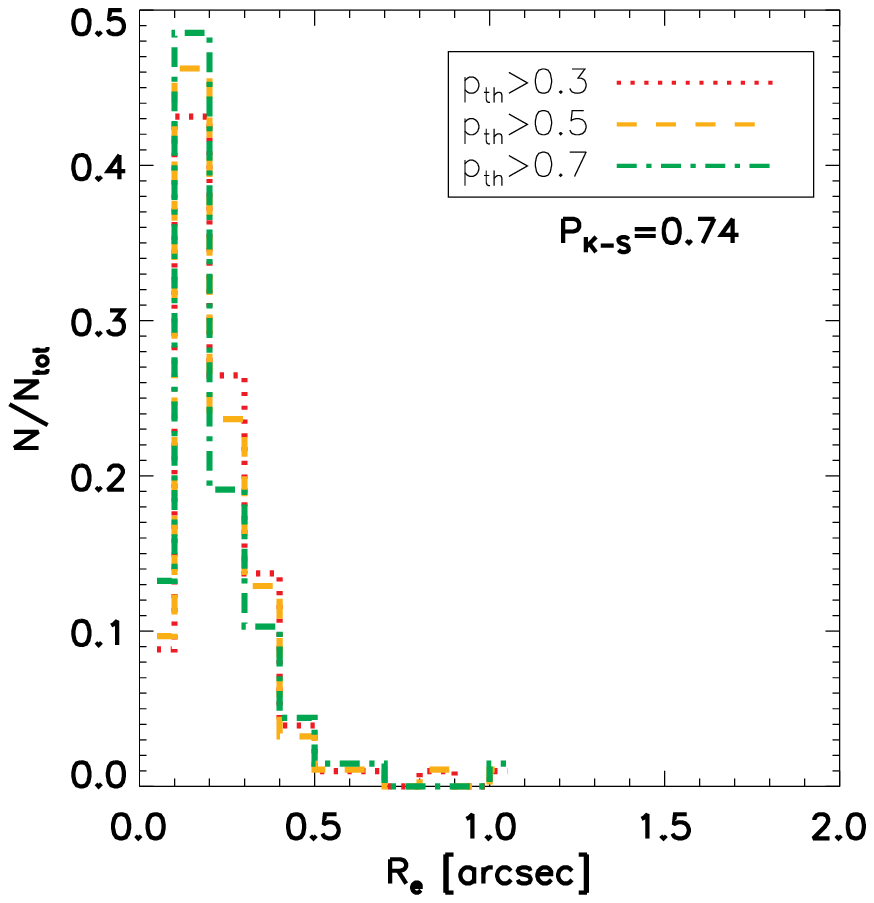} & \includegraphics[width=2.8in]{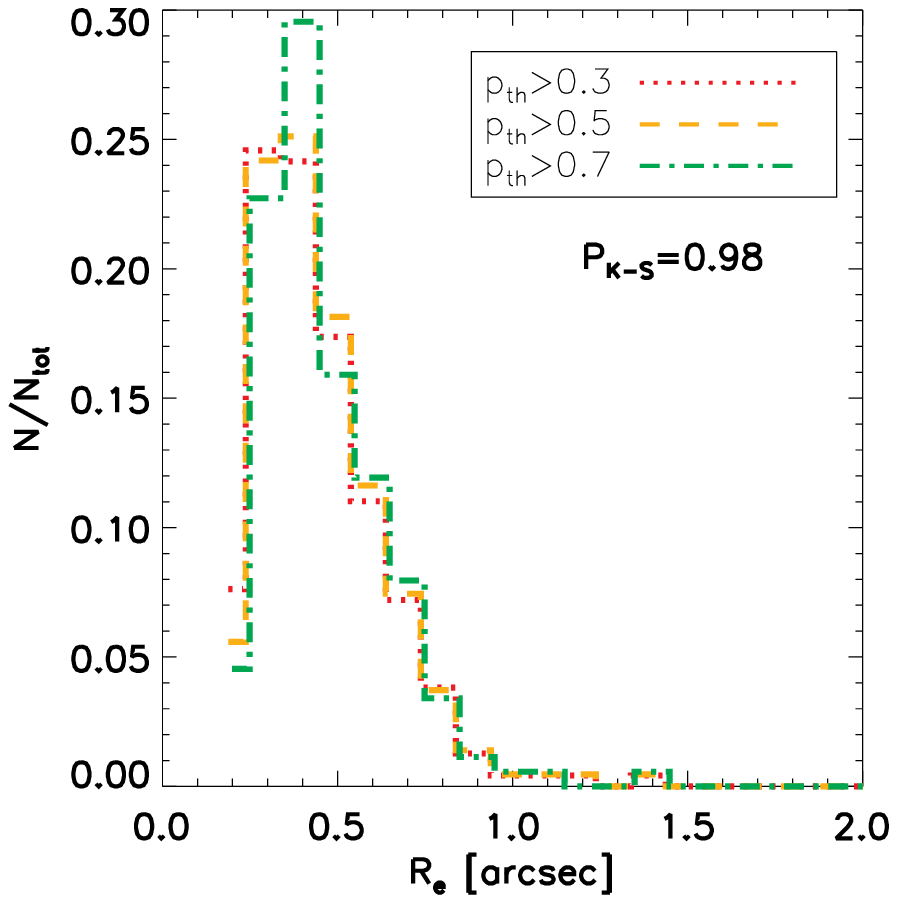}\\
\end{array}$
\caption{Surface brightness (top row), magnitude (middle row) and size (bottom row) distributions for early-type (left column) and late-type galaxies, selected with different probability thresholds. An increase in the probability threshold does not introduce significant biases in the selected galaxy population. The probability of the K-S test between the $p_{th}>0.3$ and $p_{th}>0.7$ sample is indicated in each panel. }
\label{fig:props_probas}
\end{center}
\end{figure*}

%.........................................................................................................

\subsection{Recovering finer morphological types: irregular disks/mergers}
We proceed by investigating how the automated classifications
perform in recovering finer morphological classes, in particular
mergers and irregular disks. Figure~\ref{fig:bayesian_probs_irr}
shows the galSVM-derived probability for galaxies to be irregular
or merger using both the local and high redshift training sets.
For this exercise, the local training set consists of
visually-identified mergers from Galaxy Zoo, while the
high-redshift training set was constructed by combining the
irregular disks and merger classes, both due to a lack of
statistics and also because, as shown in Section~\ref{sec:visual},
the distinction between these two classes is challenging to
produce even visually.

The left-hand panel of Figure~\ref{fig:bayesian_probs_irr} shows
that ETGs have very low galSVM probabilities of being irregular,
while visually-classified mergers and irregular disks have larger
probabilities. galSVM is, therefore, able to distinguish irregular
objects from ETGs with good accuracy. The regular LTG population,
however, does have a broader probability distribution, which is a
reflection of the fact that the general population of
late-type galaxies at $z>1$ tends to be more asymmetric than in the
local Universe, as shown in section~\ref{sec:zevol}. The accuracy
of the classification for different probability thresholds is
quantified in Table~\ref{tbl:ETG_C_P}, which shows that irregular
objects can be recovered by galSVM with a (low) contamination of
$\sim20\%$ and a (fairly high) completeness of $\sim80\%$. Some
example images of galaxies automatically classified as irregular
are shown in figure~\ref{fig:vstamps_irr_lambda}

We also explore the effect of the training set used on the
recovery of mergers and irregular disks at $z>1$. The right-hand
panel of Figure~\ref{fig:bayesian_probs_irr} shows the
galSVM-derived probability for a galaxy to be irregular, obtained
using the low-redshift (SDSS) training set. While the general
trends are preserved, the probability distribution for the merger
class is broader than the one obtained with the high redshift
training set. This is driven by the fact that the population of
mergers and irregular disks observed at high redshift do not have
direct analogues in the local Universe. Hence, the training set
lack good `templates' for these objects which are required to
train the SVM accurately. This has an obvious impact on the final
classification, especially in terms of the completeness, as can be
seen in Table~\ref{tbl:ETG_C_P}. In other words, for the same
probability threshold, the completeness is $\sim20\%$ less with a
local training sample than with the high redshift training set.

\begin{figure*}
\begin{center}
$\begin{array}{c c}
\includegraphics[width=3.0in]{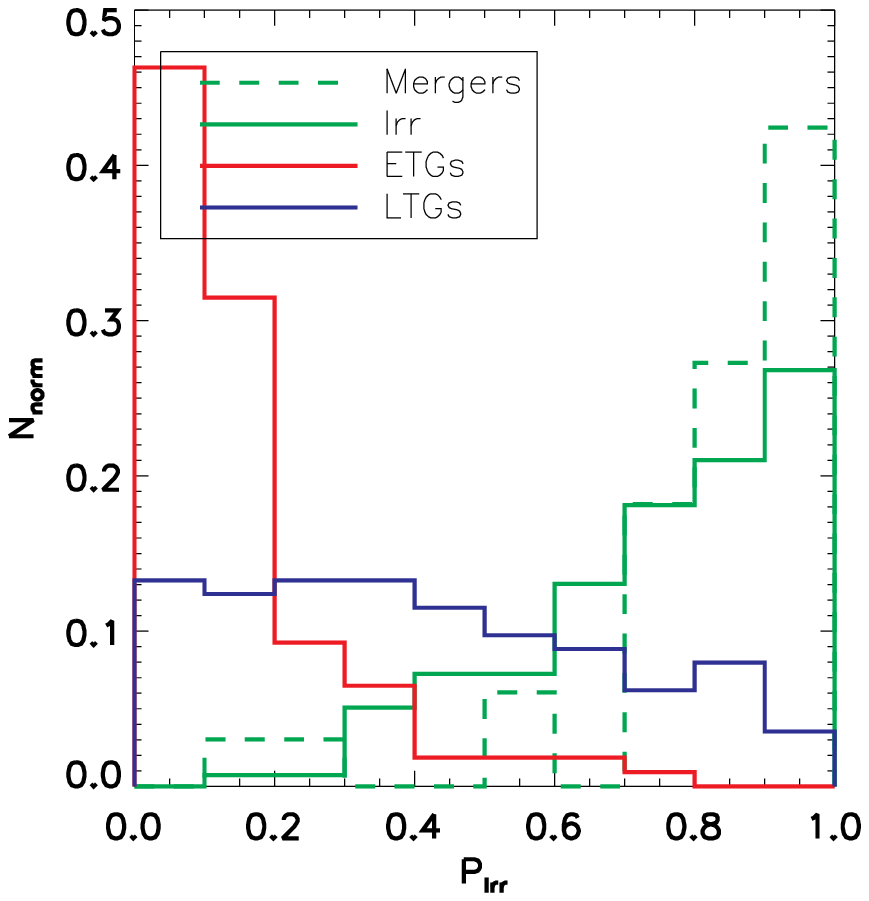} & \includegraphics[width=3.0in]{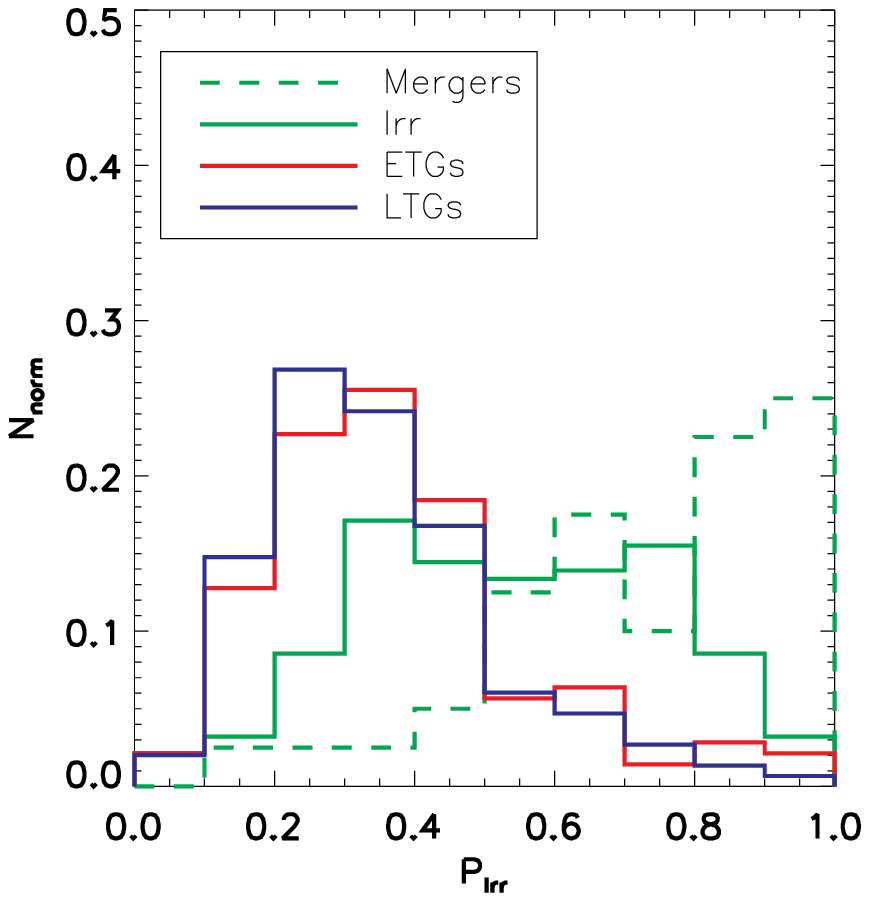}\\
\end{array}$
\caption{Irregular/merger probability distribution for visually identified morphological types as labelled. The left panel shows the result when a high z training sample is used and the right panel the result with the SDSS dataset.}
\label{fig:bayesian_probs_irr}
\end{center}
\end{figure*}

\begin{figure*}
\includegraphics[width=0.99\textwidth]{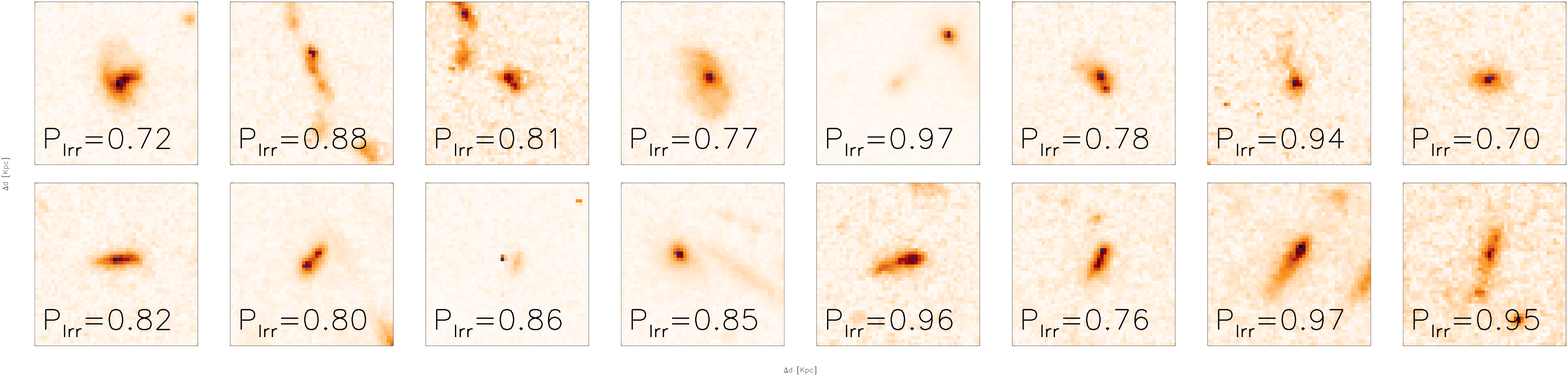}
\caption{H band cutouts of automatically classified irregular/merger galaxies selected with a probability threshold $p>0.7$ (i.e. $\sim 15\% $ contaminations). The associated probability is indicated in each individual stamp.}
\label{fig:vstamps_irr_lambda}
\end{figure*}

%.........................................................................................................

\section{Comparison with CANDELS visual classifications}
\label{sec:candels} 
Recently, the CANDELS team (PIs Faber \&
Ferguson; see Grogin et al. 2011, Koekemoer et al. 2011) has
performed  a detailed visual classification of galaxies in the
GOODS-S field \citep{2014arXiv1401.2455K} using the H-band WFC3
observations down to $H<24.5$. In this significant
effort, 65 individual classifiers contributed to the morphological classification (being 3-5 the average number of classifiers per galaxy)
 who are asked to provide a number of
flags related to the galaxy's structure, morphological
k-correction, interaction status and clumpiness. As a result,
each galaxy in the catalog has an associated fraction of
classifiers who have selected a given flag which can be used to
define morphological classes. This dataset is therefore an
excellent independent test to assess the accuracy of the automated
classification scheme presented in this work.

Following \cite{2014arXiv1401.2455K}, we first define 4 morphological
classes - mostly disks, mostly spheroids, disk + spheroid and
irregulars -  based on the fractions:
\begin{itemize} 
\item {\it Mostly disks} are those
galaxies for which less than 2/3 of the classifiers selected them
as spheroids and more than 2/3 selected them as disks. 
\item {\it Mostly
spheroids} are those for which more than 2/3 voted for a spheroid
and less than 2/3 for a disk. 
\item The \emph{disk + spheroid} class contains
galaxies that were labelled as spheroids and disk by more than 2/3
of the classifiers. 
\item Finally, the irregular class is made of
galaxies identified as irregular by more than 2/3 of the
classifiers and for which less than 2/3 selected them as disks or
spheroids. 
\end{itemize}
These classes should roughly correspond with the visual
classes selected in this work (see section~\ref{sec:visual}). Recall, however, that our visual classification above
defined an `irregular disk' class which might not be in these 4
CANDELS classes just described. In order to cope with this, we split the mostly disks class into asymmetric and regular
objects using the $f_{asym}$  fraction (lower and larger than $2/3$) which provides the fraction
of classifiers that label the galaxy as irregular.

We then perform 2 different tests. First, we simply calibrate the effect of using different visual classifications to estimate the accuracy of the classifier, by plotting the distribution of
the probabilities (‘trained’ on the ERS bases classes described in section~\ref{sec:visual}) of being labelled as ETG (left panel) and
irregular (right panel) for the ERS galaxies but for the 5 visually defined classes in
CANDELS as explained above (Figure~\ref{fig:candels_probas}). The purpose of this test is to estimate the robustness of the automated and visual classifications but we still keep a redundancy in the training and testing sets. We recover the expected trends confirming the results of previous sections. All disk
dominated galaxies (disks, irregulars and disk irregulars) present
low probabilities to be ETG while the spheroid class has high
values. Broad morphological types are therefore well recovered at
$z>1$. Interestingly, the bulge+disk class from CANDELS has large
probabilities to be ETG according to galSVM, which means that the
bulge component is dominating the computation of the morphological
parameters. Concerning the irregular galaxies, the right panel of
figure~\ref{fig:candels_probas} confirms the trends observed in
the previous sections. Galaxies visually classified as irregular
by the CANDELS team exhibit a peak at large galSVM probabilities
to be irregular. However, early-type galaxies peak at low
probabilities, which confirms the fact that irregulars can indeed
be easily separated from ETGs. As was the case for the our
visual classifications based on the ERS, the regular disk class
presents a flat distribution and will therefore pollute the
automatically selected class of irregulars. These trends are quantified in the first 2 columns of
table~\ref{tbl:ETG_C_P_CANDELS} in terms of purity and completeness, as in
previous sections. We note that the values for the three
morphological classes are comparable to the values reported using
our ERS visual classification above. The main conclusion of this test is that visual classifications are robust enough so that the accuracy of the automated classification is not significantly affected.

\begin{figure*}
\begin{center}
$\begin{array}{c c}
\includegraphics[width=3.0in]{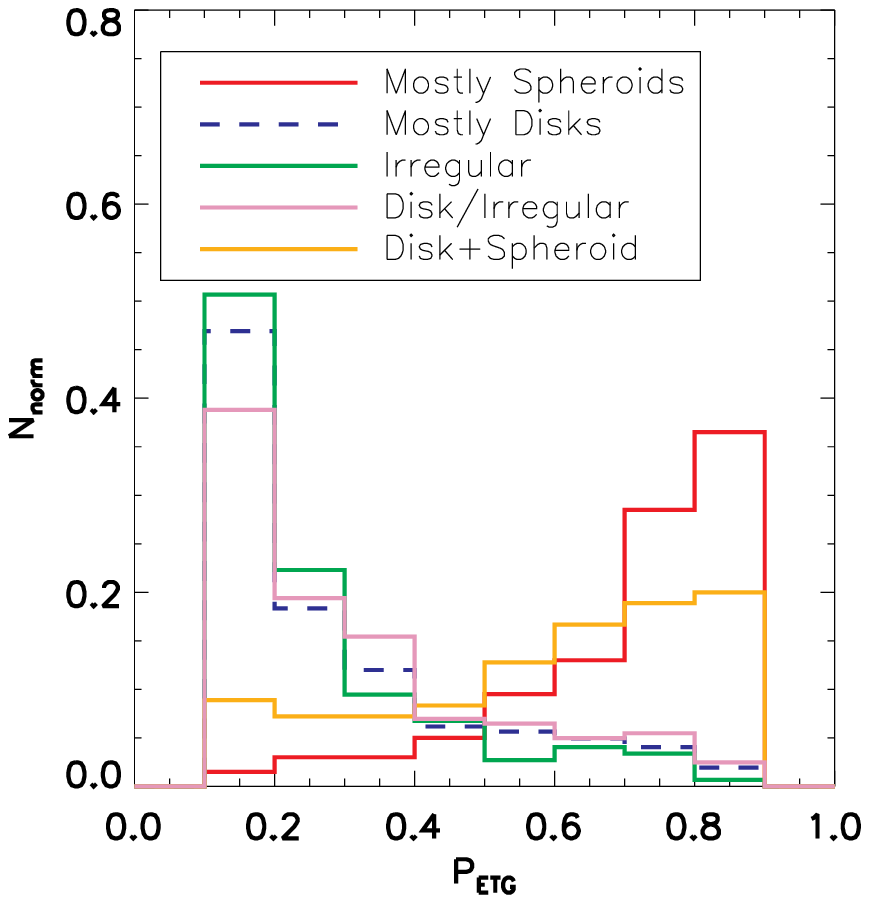} & \includegraphics[width=3.0in]{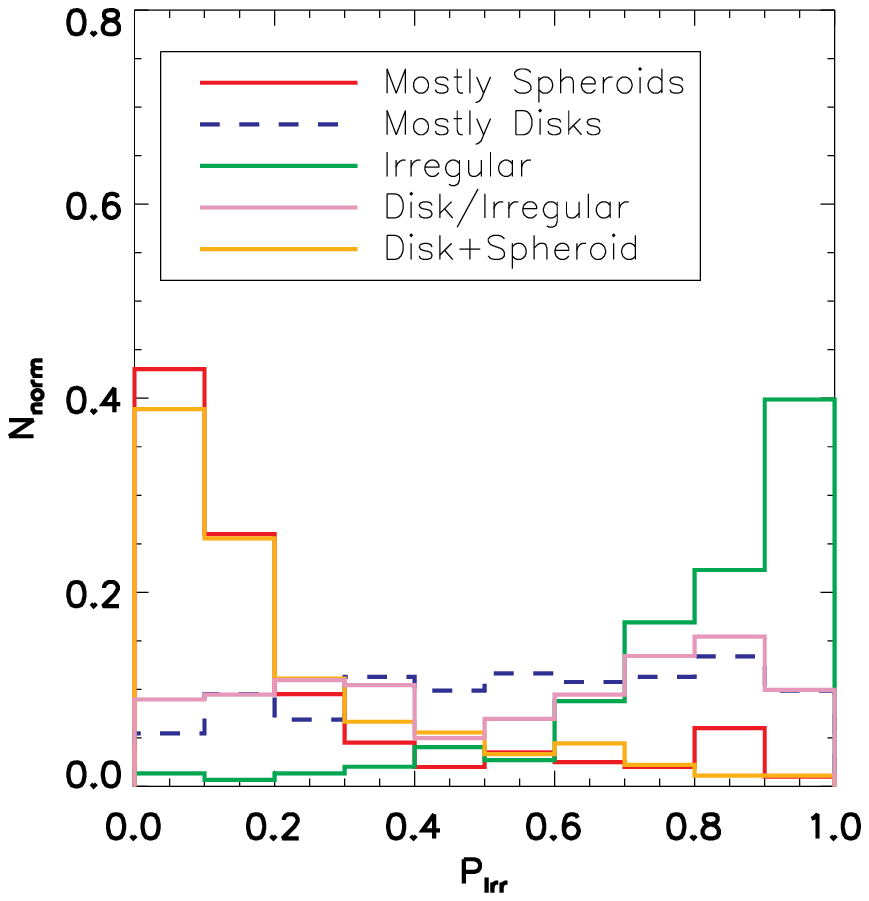}\\
\end{array}$
\caption{Comparison of the automated classification on the ERS sample with the CANDELS visual classification. Left panel: probability distribution to be ETG for different visually identified morphological types in CANDELS as labelled (see text for details). Galaxies visually classified as spheroids have large probabilities to be ETG. The Spheroid+Bulge class tends to have also high probabilities. Right panel: probability distribution to be irregular for different visually identified morphological types in CANDELS as labelled (see text for details). Galaxies visually classified as irregulars have large probabilities to be irregular according to the automated scheme presented in this work. The contamination comes mostly form the disk class (blue dashed line).}
\label{fig:candels_probas}
\end{center}
\end{figure*}

The CANDELS sample allows however a more crucial test regarding the independence of the training set. So far, we have used overlapping samples for training and classifying as explained in section~\ref{sec:auto_morph}, which necessarily leads to a \emph{best case} solution. The main purpose of automated classification schemes however is to be used massively in a large dataset for which visual classification are available only for a small fraction of the sample. We now split the full CANDELS morphological catalog in two independent subsets and use one for training and the other for testing (with the CANDELS visual classes as training labels). We select galaxies with $H<24$ and $1<z<3$ in the GOODS-S area with visual morphologies from the catalogs of \cite{2014arXiv1401.2455K} (i.e. 1717 objects). The photometric redshifts for this sample are taken from the 3D-HST public release v4.1\citep{2014arXiv1403.3689S,2012ApJS..200...13B} This test should therefore calibrate the effect of the redundancy on the training set. Results are shown in figure~\ref{fig:candels_probas_CTRAIN} and quantified in table~\ref{tbl:ETG_C_P_CANDELS}. We again recover very similar trends, (especially for the early-type class) to the ones already discussed, which proves that the classification is robust even when independent datasests are used for training and testing. Interestingly, the behavior of the irregular class slightly changes and the peak at large probabilities becomes less pronounced resulting in a smaller completeness (see table~\ref{tbl:ETG_C_P_CANDELS}). This certainly needs to be investigated more carefully in future work but it might be  a reflection of the fact that the irregular class is less well definded than other classes (in visual classifications) and hence the learning machine tends to be very training set dependent. It could also be an effect of S/N since the CANDELS are shallower on average than the ERS.

\begin{figure*}
\begin{center}
$\begin{array}{c c}
\includegraphics[width=3.0in]{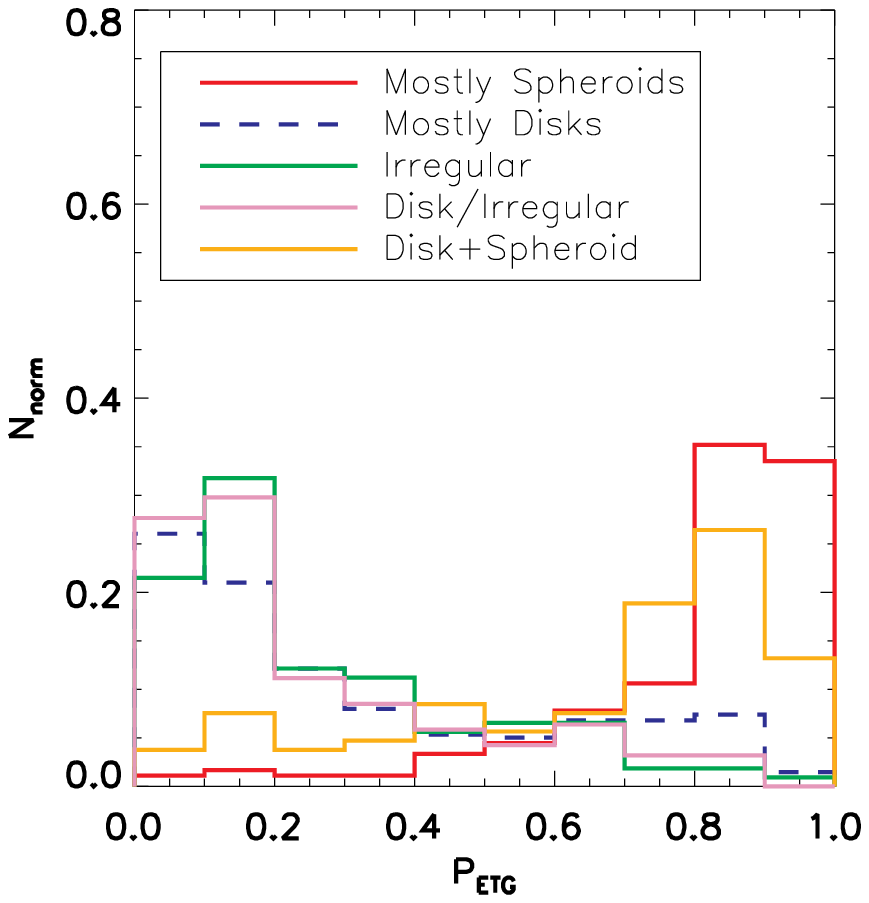} & \includegraphics[width=3.0in]{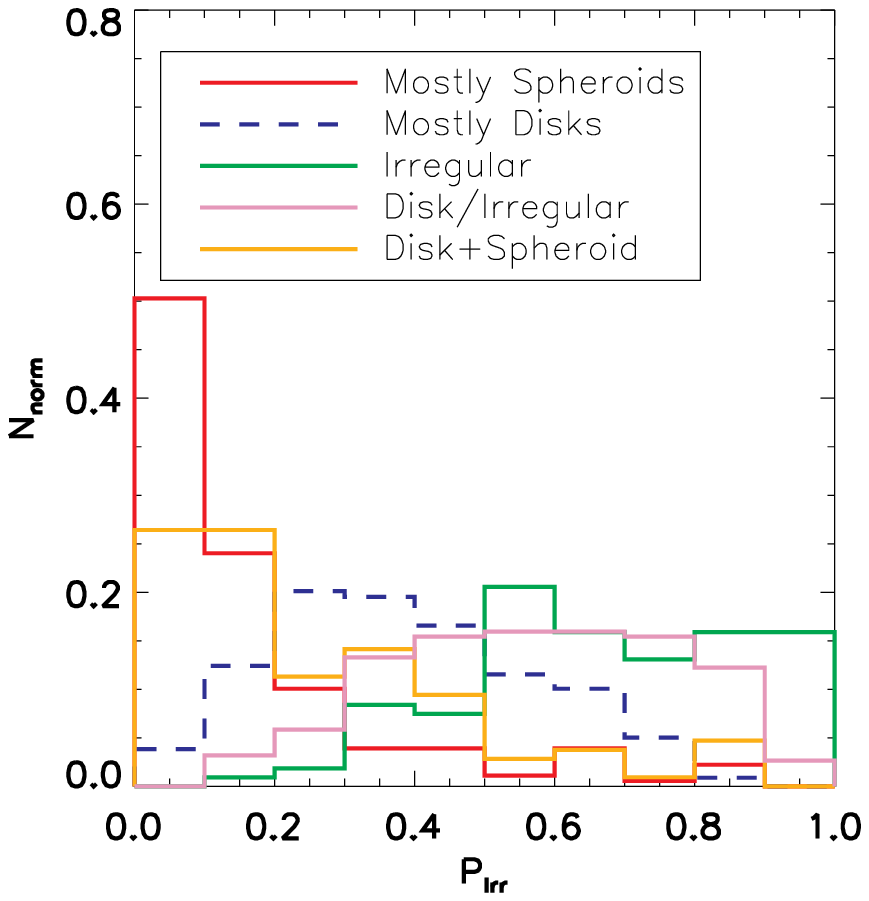}\\
\end{array}$
\caption{Comparison of CANDELS automated classification trained on the CANDELS visual classification using an independent subsample. Left panel: probability distribution to be ETG for different visually identified morphological types in CANDELS as labelled (see text for details). Galaxies visually classified as spheroids have large probabilities to be ETG. The Spheroid+Bulge class tends to have also high probabilities. Right panel: probability distribution to be irregular for different visually identified morphological types in CANDELS as labelled (see text for details). Galaxies visually classified as irregulars have large probabilities to be irregular according to the automated scheme presented in this work. The contamination comes mostly form the disk class (blue dashed line).}
\label{fig:candels_probas_CTRAIN}
\end{center}
\end{figure*}

\begin{table*}
\begin{center}
\begin{tabular}{c c c c c}

        $P_{thresh}$    & $P^{ERS}(CANDELS)$ & $C^{ERS}(CANDELS)$ & $P^{CANDELS}(CANDELS)$ & $C^{CANDELS}(CANDELS)$\\
        \hline
        \hline
        {\bf ETGs} &&&\\
        \hline
0.3 &  50.93 & 92.44  & 58.45 & 92.83\\
0.4  & 57.61 & 89.08 & 64.54 & 89.14\\
0.5  &  66.67 & 85.71 & 69.41 & 84.63\\
0.6  & 73.64 & 79.83 & 73.50 & 80.12\\
0.7  & 78.43 & 67.23 & 79.55 & 72.54\\
0.8  & 91.53 & 45.38 & 86.39 & 59.84\\
\hline
 {\bf LTGs} &&&\\
 \hline
 0.3 & 88.39 & 93.10 & 86.75 & 90.60\\
0.4  & 92.23 & 89.34 & 89.50 & 85.43\\
0.5  & 94.04 & 84.01  & 91.29 & 81.20\\
0.6 & 94.88 & 75.55 & 93.22 & 75.31\\
0.7 & 95.95 & 66.77 & 94.86 & 66.74\\
0.8 & 96.89 & 48.90 & 95.88 & 55.27\\
\hline
 {\bf Irr/mergers} &&&\\
 \hline
0.3  &55.63 & 90.80 & 47.05 & 94.21\\
0.4 & 58.57 & 84.48 & 52.97 & 85.30\\
0.5 & 62.96 & 78.16 & 60.73 & 73.72\\
0.6 & 67.21 & 70.69 & 67.25 & 59.47\\
0.7  & 72.00 & 62.07 & 77.47 & 43.65\\
0.8  & 73.15 & 45.40 & 83.85 & 24.28\\
\end{tabular}
\caption{Purity (P) and Completeness (C) for ETGs, LTGs and irregulars computed with the CANDELS visual classifications (Kaltartepe et al. 2014) with an ERS training set (first 2 columns) and a CANDELS independent dataset (last 2 columns) (see text for details). The superscript indicates the training set used and the name in the brackets the classified sample}
\label{tbl:ETG_C_P_CANDELS}
\end{center}
\end{table*}

%.........................................................................................................

\section{Multi-wavelength classification}
\label{sec:multi_lambda} In this section we explore how the
performance of the automated classifications changes when we use
information in different wavelengths.

\subsection{UV vs. optical} At $z>1$, the NIR filters on the HST
trace the rest-frame optical wavelengths, while the optical
filters probe the rest-frame UV. In Section~\ref{sec:uv_opt} we
showed that there are moderate variations in the values of our
morphological parameters depending on the wavelength being
employed to measure their values. We now quantify how the galSVM
classifications respond to a change in wavelength.
Figure~\ref{fig:ph_pi} presents the galSVM-derived probabilities
of galaxies to be classified as ETG and irregular, obtained for
the same objects with parameters measured in the $H$ and $i$ bands
(which traces the rest-frame optical and UV respectively). The
training set is the same for both classifications, i.e. visual
classifications performed in the $H$ band using the ERS.

We find a good correlation between the galSVM-derived
probabilities based on different wavelengths. Galaxies with high
probabilities to be ETG in the $H$ band generally have high
probabilities to be ETG in the $i$ band. In fact, the best fit
line between the two probabilities gives almost a one to one
relation, i.e. $p_{ETG}^i=(1.1\pm0.04)\times
p_{ETG}^{H}-(0.05\pm0.01)$ and the scatter of $\sim0.2$. A similar
behaviour is observed for irregular galaxies but the best fine
line has a slope $<1$, i.e. $p_{irr}^i=(0.73\pm0.02)\times
p_{irr}^{H}+0.13\pm0.01)$. There are however a few ($<2\%$)
\emph{outliers} which present high (low) probabilities to be ETG
in the $H$ band and high (low) in the $i$ band. The differences in
the classifications appear not to be a consequence of the
difference in the shape of the two filters. As a matter of fact,
the values of the morphological parameters used for the
classification are very similar in both filters most of the times.
The discrepancy in the final classification would be better
explained by a failure in the algorithm, i.e outliers, that will
be carefully explored in future work.

\subsection{H+i classification}
We now explore the effect of \emph{simultaneously} using
morphological parameters measured in the $H$ and $i$ bands on the
galSVM classifications. Table~\ref{tbl:ETG_C_P_multi} summarizes
the purity and completeness of early and late-type galaxy samples
obtained. Interestingly, we do not detect a significant change in
the accuracy of the final classification when these additional
parameters are included. Since, as shown in
Section~\ref{sec:uv_opt}, there is a strong correlation between
parameters measured in the two filters, using parameters measured
in both bands does not add enough information to increase the
quality of the classification. We obtain similar results when more
than two filters are used. This, however, indicates the robustness
of SVM to a redundancy in the data since, despite the inclusion of
highly redundant parameters, the accuracy of the classification
remains unaltered.

\subsection{Adding additional \emph{no-morphological} information: stellar mass} We conclude
this section by exploring if the accuracy of the final
classification can be improved by adding additional information derived
from the spectral energy distribution of the galaxies, in this
case the stellar mass. We re-classify the full sample using the
main morphological parameters measured in the $H$ band adding
stellar mass as an extra parameter keeping the same training set
based exclusively on the visual aspect of the galaxy. In other
words, the information brought by the stellar mass is used a
posteriori, to check if better accuracy can be reached. The
results in terms of purity and completeness are summarized in
Table~\ref{tbl:ETG_C_P_mstar}, for early and late-type galaxies.
The inclusion of the stellar mass in the classification process
tends to increase the completeness of the final classification,
especially for the early-type population. For an equivalent
probability threshold, e.g. P$\sim$0.6, a classification which
uses the stellar mass is $\sim$10\% more complete.

\begin{figure*}
\begin{center}
$\begin{array}{c c}
\includegraphics[width=3.0in]{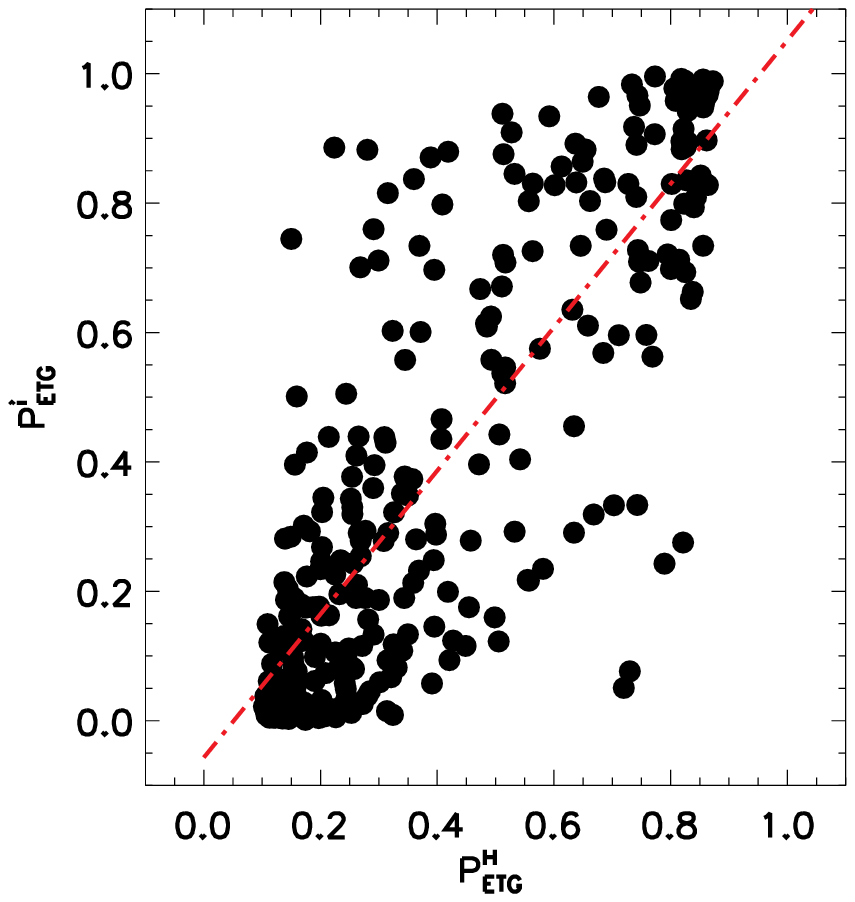} & \includegraphics[width=3.0in]{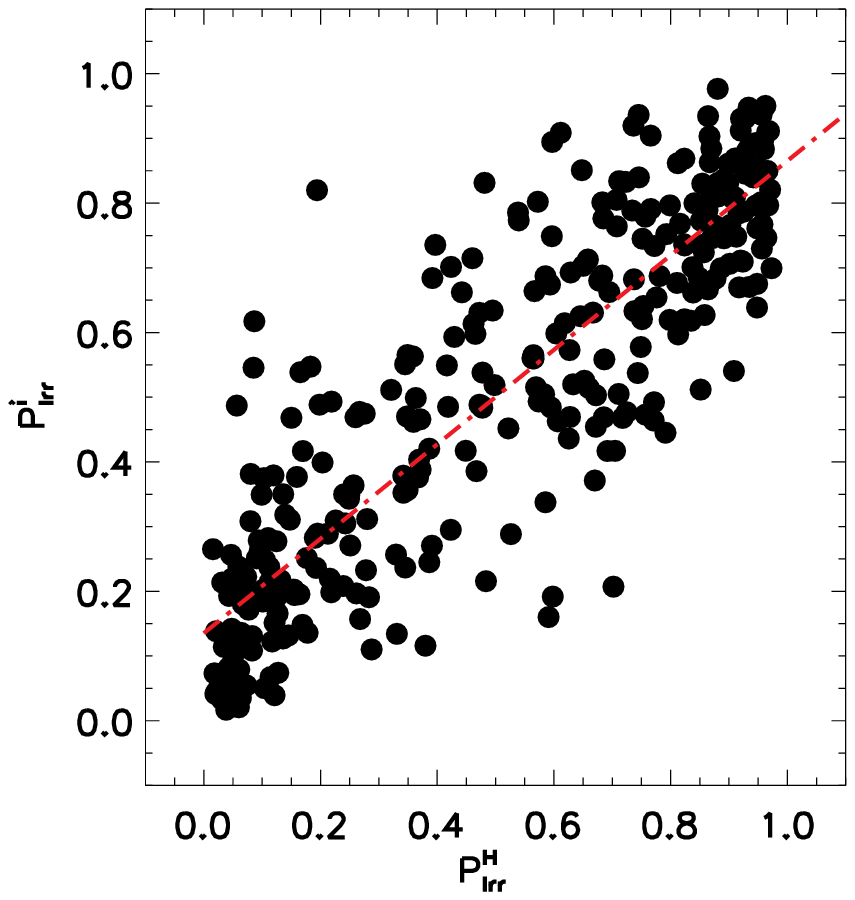}\\

\end{array}$
\caption{Probability to be ETG derived in the H band  vs. the same probability derived in the i band (left panel). The right panel shows the probabilities to be irregular derived in the H and i band. The red dashed-dotted line shows the best fit line.}
\label{fig:ph_pi}
\end{center}
\end{figure*}

\begin{table*}
\begin{center}
\begin{tabular}{c c c}

        $P_{thresh}$           &  P & C   \\
        \hline
        \hline
{\bf ETGs}&&\\
\hline
0.3  & 49.07 & 96.34\\
0.4  & 57.14 & 92.68\\
0.5 & 63.79 & 90.24\\
0.6  & 70.71 & 85.37\\
0.7  & 78.48 & 75.61\\
0.8  & 80.30 & 64.63\\
\hline
{\bf LTGs}&&\\
\hline
0.3  & 92.00 & 93.12\\
0.4 & 94.78 & 88.26\\
0.5  & 96.24 & 83.00\\
0.6 & 96.94 & 76.92\\
0.7 & 98.21 & 66.80\\
0.8 & 99.25 & 53.85\\
\end{tabular}
\caption{Purity (P) and Completeness (C) for ETGs classified using parameters measured in the H and i band simultaneously (see text for details)}
\label{tbl:ETG_C_P_multi}
\end{center}
\end{table*}

\begin{table*}
\begin{center}
\begin{tabular}{c c c}

        $P_{thresh}$            & P  & C   \\
        \hline
        \hline
{\bf ETGs}&&\\
\hline
0.3 & 70.69 & 98.80\\
0.4 & 76.70 & 95.18\\
0.5 & 81.25 & 93.98\\
0.6 & 84.88 & 87.95\\
0.7  & 86.25 & 83.13\\
0.8 & 86.49 & 77.11\\

        \hline
{\bf LTGs}&&\\
\hline
0.3 & 91.62 & 93.29\\
0.4  & 93.79 & 92.07\\
0.5 & 96.69 & 89.02\\
0.6  & 97.22 & 85.37\\
0.7 & 99.24 & 79.27\\
\end{tabular}
\caption{Completeness and purity for ETGs classified using parameters measured in the H and the stellar mass simultaneously (see text for details)}
\label{tbl:ETG_C_P_mstar}
\end{center}
\end{table*}

%.........................................................................................................

\section{Summary}
\label{sec:summary} We have studied the morphologies of 628 bright
($H<24$ mag) galaxies in the redshift range ($1<z<3$), using both
visual inspection of images and commonly-used morphological
parameters - Concentration ($C$), Asymmetry ($A$), the Gini
coefficient ($G$) and $M_20$.

The overall aim of this paper has been to quantify how different
morphological classes evolve structurally over 90\% of cosmic time
and how well morphological parameters recover these classes at
$z>1$, compared to visual inspection.

%We also provide a morphological classification of CANDELS-GOODS.... (think about that. It's almost done and it might maximize the impact....)

\begin{enumerate}

\item In the first part of this paper, we have studied the general
behaviour of usual morphological proxies (i.e. C, A, S, G,
$M_{20}$)  at $z>1$, as well as their wavelength dependence and
redshift evolution:

\begin{itemize}

\item Generally speaking, the behaviour of morphological
parameters at $z>1$ is similar to that at low redshift, in the
sense that bulge-dominated galaxies are more concentrated and less
asymmetric than disk dominated galaxies, with mergers and
irregular disks being highly asymmetric.

\item By artificially redshifting local galaxies, we have measured
the evolution of the morphological parameters with redshift. All
parameters except asymmetry show moderate redshift evolution. At
$z>1$, galaxies are $>50\%$ more asymmetric, have $\sim10\%$
larger Gini coefficients and 10\% lower M20 values than at $z=0$. The
evolution is more pronounced for the early-type population.

\item There is a good correlation between parameters measured in
the rest-frame optical and UV wavelengths, especially for ETGs.
Late-type galaxies tend to be more asymmetric and more clumpy when
observed in the UV rest-frame by $40-50\%$.

\end{itemize}

\item In the second part of this paper, we have explored the
ability of automated parameters to recover morphological classes.
We have applied the support vector machine based code
\textsc{galSVM} to the full sample, using all the morphological
parameters simultaneously and two different training sets (one
local and one at high redshift) and have derived probabilities of
being ETG, LTG and Irregular/merger for individual galaxies.

\begin{itemize}

\item Early-type galaxies can be recovered automatically with a
contamination of $< 20\%$ and a completeness of $\sim 80\%$. For
LTGs, less than $5\%$ contamination is measured with a
completeness of $90\%$ depending on the probability threshold
applied.

\item Irregular classes and mergers can be recovered with a purity
of $\sim 70\%$. The contamination mainly derives from disk
galaxies which are typically more irregular at $z>1$ than in the
local Universe.

\item The accuracy of the classification decreases when a local
training set based on the SDSS is used. This is a consequence of
the evolution of the morphological parameters. This is especially
critical for the irregular classes.

\item The automatic classification presented in this work correlates well with the CANDELS visual classificiation. Similar values of completeness and purity as the ones shown above are obtained when the classification is either compared to the CANDELS morphologies keeping the same training set or when the classifier is trained using the CANDELS classifications. 

\item There is a good correlation between the automated classification
performed in the rest-frame optical and UV wavelengths with less than 2\% galaxies presenting very different probabilities in the two filters.

\item Adding the stellar mass derived from the SED to the classifier such as the stellar mass and/or morphological parameters measured in different filters, tends to increase the slightly increase the completeness of the classification. A more detailed analysis will be performed in future work.

\end{itemize}

Given the prodigious amount of data expected from future surveys
such as EUCLID and LSST, a unified approach that combines both
visual inspection and morphological parameters is desirable.
Accurate visual inspections are necessary to properly train the
algorithms and maximize the efficiency of the automatic
procedures. However, we have shown that automated techniques,
based on machine learning do a reasonable job in recovering
morphological classes at $z>1$, including fine morphological
classes (e.g. irregular/mergers). In forthcoming papers, we will provide
a classification of all CANDELS fields with the procedure
presented here. We will also explore new methodologies based on novel machine learning algorithms.

\end{enumerate}

\begin{comment}
\begin{figure}
\begin{center}
$\begin{array}{c}
\includegraphics[width=3.3in]{histograms_visual_class_petg_ETGs.ps}\\
\includegraphics[width=3.3in]{histograms_visual_class_petg_LTGs.ps}\\
%\includegraphics[width=3.3in]{P_bulge.ps}
\end{array}$
\caption{Eary-type and late-type probability distributions.} \label{fig:bayesian_probs}
\end{center}
\end{figure}
\end{comment}

%.........................................................................................................

\section*{Acknowledgements}
S.K. acknowledges fellowships from the 1851 Royal Commission,
Imperial College London, Worcester College Oxford and support from
the BIPAC Institute at Oxford. M.H.C would like to thank Jeyhan Kartaltepe for kindly providing us with the CANDELS morphology catalog. This paper is based on Early Release Science observations made by the WFC3 Scientific Oversight Committee. We thank the Director of the Space Telescope Science Institute for awarding Director’s Discretionary time for this program. Support for program 11359 was provided by NASA through a grant from the Space Telescope Science Institute, which is operated by the Association of Universities for Research Inc., under NASA contract NAS 5-26555. This work is based on observations taken by the 3D-HST Treasury Program (GO 12177 and 1232) with the NASA/ESA HST, which is operated by the Association of Universities for Research in Astronomy, Inc, under NASA contract NAS5-26555.

%.........................................................................................................

\nocite{Giavalisco2004,Dekel2009a,Dekel2009b,Keres2009,Kimm2012,Kaviraj2012,Windhorst2011}

\bibliographystyle{mn2e}
\bibliography{references}

%.........................................................................................................

\end{document}